\documentclass[twocolumn,showpacs,prb]{revtex4-1}%
\usepackage{amsmath}
\usepackage{amssymb}
\usepackage{amsfonts}
\usepackage{bm}
\usepackage{amssymb}
\usepackage[dvips]{graphicx}
\usepackage{dcolumn}
\usepackage{txfonts}
\usepackage{hyperref}
\usepackage{makeidx}
\usepackage{color}
\usepackage{mathtools}

\begin{document}

\title{Decoherence and dynamical decoupling control of nitrogen-vacancy center electron
spins in nuclear spin baths}
\author{Nan Zhao}
\affiliation{Department of Physics and Centre for Quantum Coherence, The Chinese University of Hong Kong, Shatin,
New Territories, Hong Kong, China}
\author{Sai-Wah Ho}
\affiliation{Department of Physics and Centre for Quantum Coherence, The Chinese University of Hong Kong, Shatin,
New Territories, Hong Kong, China}
\author{Ren-Bao Liu}
\email{rbliu@cuhk.edu.hk}
\affiliation{Department of Physics and Centre for Quantum Coherence, The Chinese University of Hong Kong, Shatin,
New Territories, Hong Kong, China}


\begin{abstract}
We theoretically study the decoherence and the dynamical decoupling control of nitrogen-vacancy center electron spins in high-purity diamond, 
where the hyperfine interaction with $^{13}$C nuclear spins is the dominating decoherence mechanism. 
The decoherence is formulated as the entanglement between the electron spin and the nuclear spins, 
which is induced by nuclear spin bath evolution conditioned on the electron spin state.  
The nuclear spin bath evolution is driven by elementary processes such as single spin precession and pairwise
flip-flops. The importance of different elementary processes in the decoherence depends on the strength of the
external magnetic field.
\end{abstract}

\maketitle

\section{Introduction}
\label{Sect:Intro}
The optically detected magnetic resonance of single nitrogen-vacancy (NV) centers in diamond \cite{Gruber1997}
triggered the research on spin coherence of single NV centers for applications in quantum information processing
\cite{Wrachtrup2001,Wrachtrup2006,Childress2006a,Dutt2007}
and ultra-sensitive metrology.
\cite{Maze2008,Balasubramanian2008,Taylor2008,Hall2010,Hall2009,Zhao2011a,Dolde2011,DeLange2011}
As one of the most important advantages, NV center electron spins have long coherence time up to the millisecond
timescale \cite{Kennedy2003,Gaebel2006,Balasubramanian2009} even at room temperature. 
The long coherence time in such systems makes them an ideal platform for studying quantum sciences and technologies.
Understanding the decoherence mechanisms and prolonging the coherence time of NV center spins are of fundamental
importance in exploring a number of surprising physical effects\cite{Zhao2011b,Huang2011arXiv} and novel applications
.\cite{Zhao2011a,Dolde2011}

\begin{figure}[tb]
  \includegraphics[width=0.8 \columnwidth]{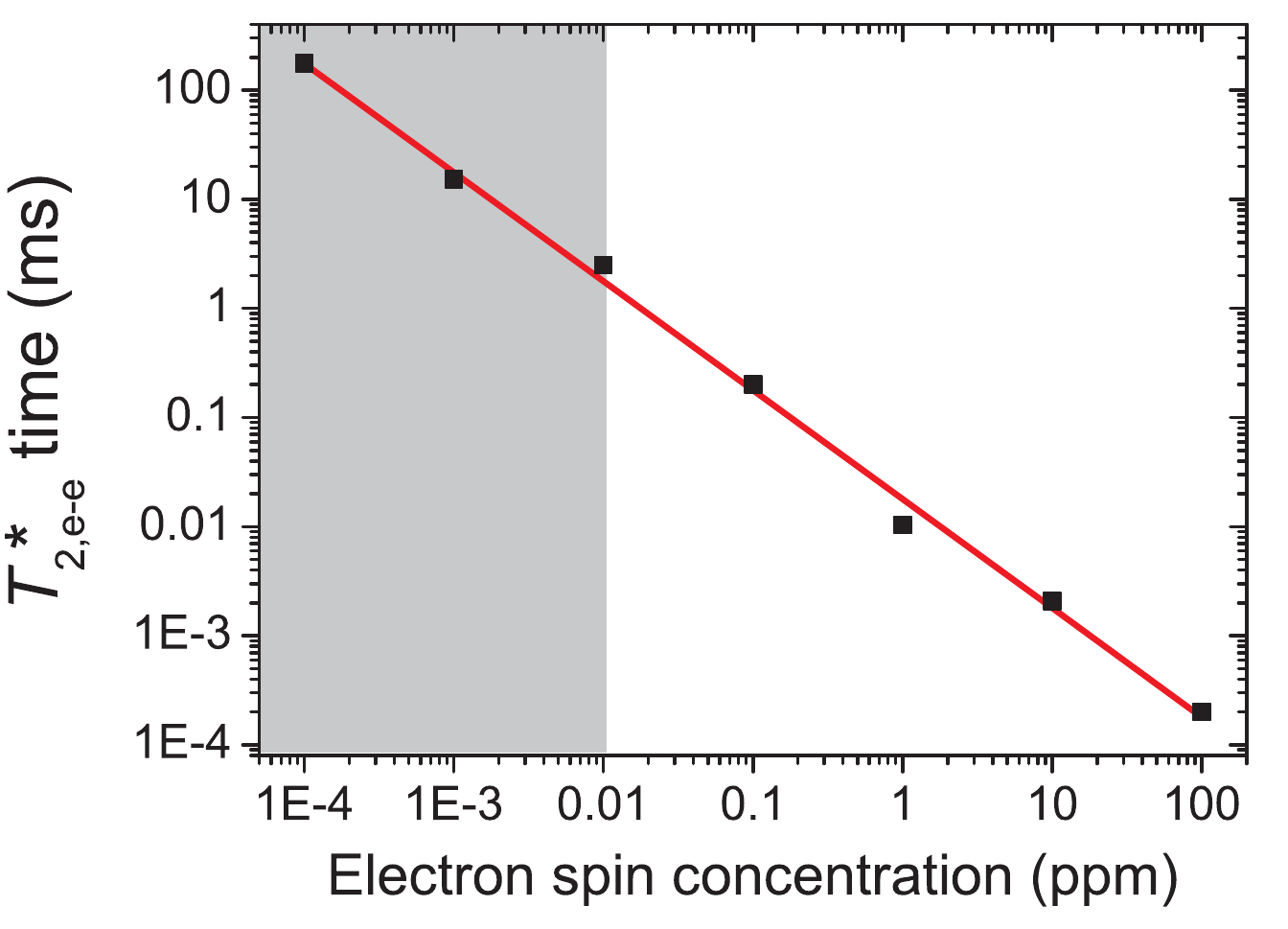}
  \caption{NV center electron spin decoherence in an electron spin bath. 
  With only the electron spin bath considered, the FID decay time $T_{2,\text{e-e}}^{*}$ is inversely proportional to
  the concentration of the bath electron spins, and exceeds $\sim 1$~ms for the concentration
  lower than 0.01~ppm (the shadowed region).}
\label{FIG:fig0_ebath}
\end{figure}

\begin{figure}[tb]
  \includegraphics[width= \columnwidth]{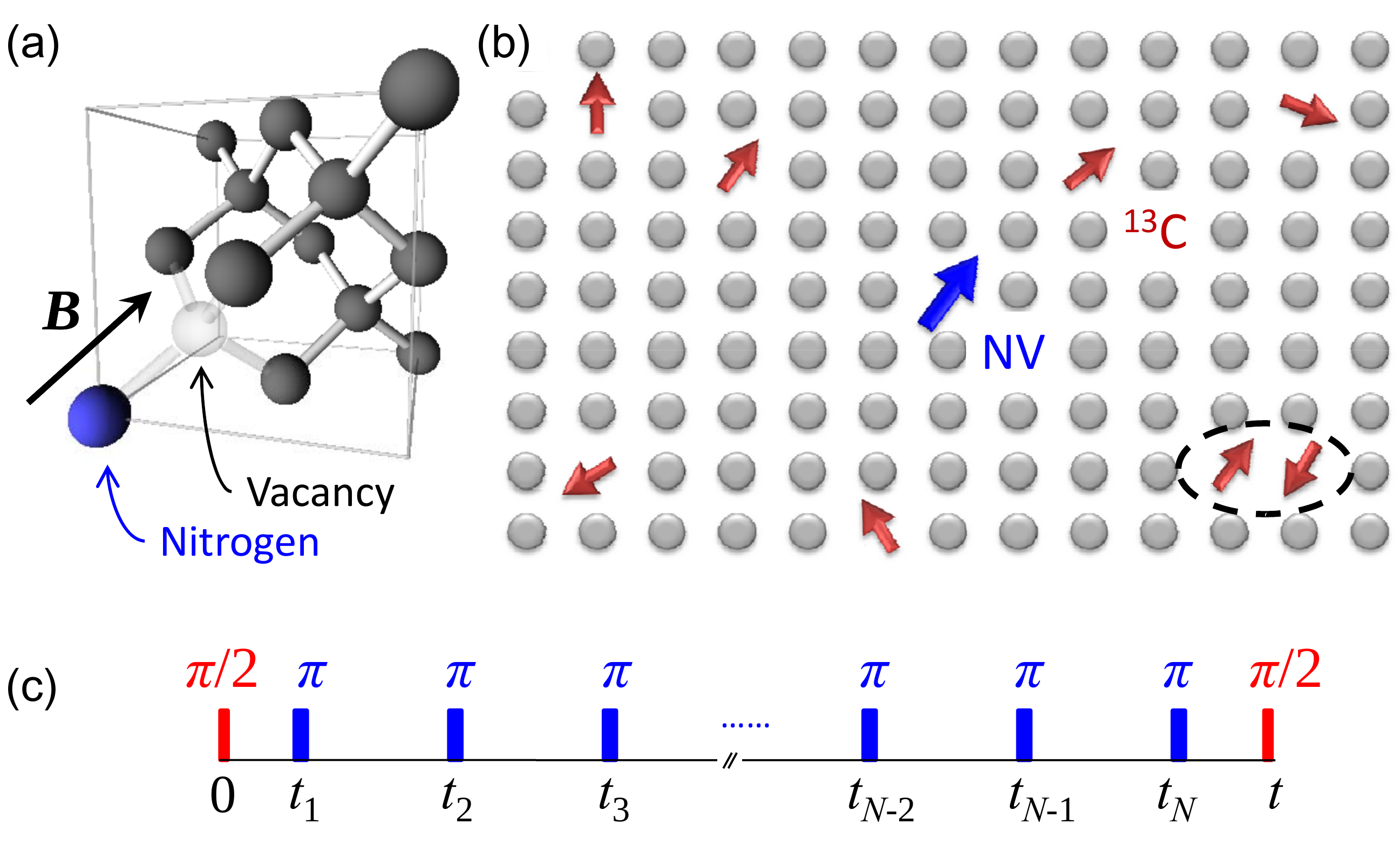}
  \caption{(a) Schematic of the NV center structure. The magnetic field is assumed to be applied along
  the direction pointing from the nitrogen to the vacancy (the $[111]$ direction). (b) Schematic of the NV center in a
  $^{13}$C nuclear spin bath. The $^{13}$C atoms with nuclear spins (the red arrows) are randomly located on the diamond lattice (the gray circles).
  A nuclear spin dimer is encircled. 
  (c) Schematic of the DD control sequence. The coherence is generated by the initial $\pi/2$ pulse (red squares). The
  electron spin is repeatedly flipped at times $t_k$ by the $\pi$ pulses (blue squares). 
  Finally, the coherence is converted to the level population by the last $\pi/2$ pulse.}
\label{FIG:fig1_schem}
\end{figure}

In general, electron spin decoherence in solids can result from phonon scattering through spin-orbit coupling,
interaction with paramagnetic impurities, and hyperfine interaction with nuclear spins.\cite{Du2009} 
Being a light-element material, diamond has weak spin-orbit interaction, and therefore the spin-lattice relaxation of NV
center spins caused by phonon scattering has rather long timescales, $T_1=7.7$~ms at room temperature and longer than seconds at low
temperature.\cite{Takahashi2008} In most cases, the NV center electron spin decoherence is not limited by the spin-lattice relaxation. 
The coupling to electron or nuclear spins in diamond can induce faster decoherence. 
In type-Ib diamond samples, the main paramagnetic centers coexisting with the NV center are nitrogen donors with one
unpaired electron spin (the P1 centers). These paramagnetic centers couple to the NV center electron spins through the
dipolar interaction, and form an electron spin bath. 
The NV center electron spin coherence time is inversely proportional to the concentration of
the P1 centers.\cite{Kennedy2003,Du2009,Witzel2010} For a P1 concentration $\sim 10^{2}$~ppm, the
NV center spin coherence time is $\sim \rm {\mu s}$.\cite{Hanson2008,Lange2010} 
In Fig.~\ref{FIG:fig0_ebath}, we plot the estimated NV center free-induction decay (FID)
time $T_{2,\text{e-e}}^{*}$ as a function of the P1 concentration, which is the lower bound decoherence time contributed
by the electron spin bath. 
When the P1 concentration is decreased to $\lesssim 10$~ppb, as in the high-purity type-IIa samples, the decay time
$T_{2,\text{e-e}}^{*}$ due to the electron spin bath will exceed millisecond. 
In this case, for NV centers in diamond with natural abundance of $^{13}\text{C}$, the decoherence will be
dominated by the hyperfine interaction with the $^{13}\text{C}$ nuclear spins, 
which form a nuclear spin bath [see Fig.~\ref{FIG:fig1_schem}(b)]. 
In this paper, we will focus on the NV center electron spin decoherence in the $^{13}\text{C}$ nuclear spin bath.

In order to protect the electron spin coherence from the environmental noises, dynamical decoupling
(DD)\cite{Viola1999,Yang2010c} control has been demonstrated as an efficient
way.\cite{Du2009,Lange2010,Ryan2010,Naydenov2011} 
In general, under DD control, the electron spins are repeatedly flipped, so that the
effect of the noises is averaged out, and the coherence time is prolonged.
In this paper, the NV center spin coherence under DD control is investigated, and
the basic physical processes in the nuclear spin bath are analyzed to reveal the decoherence mechanisms in different
magnetic field regimes.

Similar problems of electron spin decoherence in nuclear spin baths have been investigated in other systems
such as electron spins in quantum dots (QDs)\cite{Yao2006,Yao2007,Liu2007} and electron spins of shallow donors in
silicon, such as Si:P\cite{DeSousa2003,Witzel2005,Witzel2007,Witzel2010} and
Si:Bi.\cite{Bismuth2010a,Bismuth2010b,Bismuth2010c}
The spin decoherence in diamond has several distinct features as compared with that in QDs and shallow donors.

First, NV centers are deep-level defects in diamond.
In QDs and shallow donors, the electron wave functions extend to a few or even tens of nanometers, and the couplings
between the electron spins and nuclear spins are dominated by the isotropic Fermi contact interaction.
In NV centers, the electron wave function is localized around the vacancy site, extending only several
angstroms. The Fermi contact part vanishes quickly as the distance increases.  
Also, the natural abundance of the $^{13}$C, the only C isotope with non-zero spin, is only about $1.1\%$.
The electron spin couples to the bath $^{13}\text{C}$ spins mainly through the dipole-dipole interaction.
The Rabi oscillation of central spins in dipolar-coupled spin baths was studied, and complex decay behavior of central
spins under continuous driving was revealed in Ref.~\onlinecite{Dobrovitski2009}.
Here we will focus on the spin decoherence behavior under pulsed-DD control. 
The anisotropic nature of the hyperfine interaction plays an important role in the NV center decoherence.

Second, the bath spins involved in the decoherence of NV centers are much fewer than those in QDs and shallow donors. 
In QDs and shallow donors, a large number of nuclear spins (e.g., $\sim10^4$ for Si:P with natural abundance
$^{29}\text{Si}$, and $10^4\sim10^6$ for QDs depending on the QD size) contribute to the decoherence. The contribution of each nuclear
spin (or spin pair) to the total decoherence is small and can be well described by the nuclear spin dynamics in the short time limit. 
In contrast, the nuclear spin baths of NV centers in diamond consist only hundreds of $^{13}$C
nuclear spins\cite{Maze2008a,Zhao2011a}, which gives rise to a long coherence time violating the short time
condition for the nuclear spin dynamics. This brings the unique features in the NV center electron spin
decoherence.\cite{Zhao2011a,Zhao2011b}

Particularly, due to the random location of the $^{13}$C nuclear spins on the diamond lattice sites, the nuclear spin
baths of NV centers could be \textit{inhomogeneous}, 
i.e., the distance from a certain $^{13}$C spin to its neighboring spin can have a large fluctuation around the
average distance between neighboring spins. 
This distance fluctuation enables closely bonded nuclear spin clusters to appear around the
NV centers. 
Due to the inverse-cubic dependence of the dipolar interaction on the inter-nucleus distance, such
closely bonded clusters could have much faster dynamics than those weakly bonded nuclear spins, and therefore be
distinguished in the decoherence process. 
As an example, in Ref.~\onlinecite{Zhao2011a}, we have demonstrated that a 
nuclear spin \textit{dimer}, i.e. two nuclear spins occupying a C-C bond, can induce characteristic oscillations
on the decoherence profile. 
Here, we will further show that the quantum dynamics of closely bonded nuclear spin clusters [see
Fig.~\ref{FIG:fig1_schem}(b)], as localized elementary excitations in the interacting nuclear spin bath, will become the
main deocoherence mechanism under certain magnetic fields.

Theoretically, to investigate the electron spin decoherence in an interacting nuclear spin bath, 
one needs to solve the quantum many-body dynamics of the spin bath.
Various methods have been developed to treat this kind of problem, 
including the cluster expansion,\cite{Witzel2007}
the pair-correlation approximation,\cite{Yao2007,Liu2007}
and the disjoint clusters approach.\cite{Maze2008a} 
When investigating electron spin coherence of NV centers under DD controls, one usually
has to take into account the higher order correlations and the dynamics beyond the short time limit. 
In this case, the cluster-correlation expansion (CCE) method\cite{Yang2008,Yang2009} previously developed in our group
is especially suitable, which has been proved to be an efficient method in treating the decoherence in correlated nuclear spin baths of finite
size.\cite{Du2009} This paper focuses on the decoherence effects of different elementary processes in the nuclear spin
baths under various conditions. Comparison of different theoretical methods is beyond the scope of this paper.

The paper is organized in the following way.
Section \ref{Sect:sys_method} gives a description of the NV center electron spin decoherence problem and the CCE method. 
In Section \ref{Sect:spin_process_pic}, we analyze the basic spin processes involved in the decoherence. 
The results of NV center electron spin coherence in different magnetic regimes under DD controls are discussed 
in Section \ref{Sect:res_disc}. 
Section \ref{Sect:conclusion} gives the conclusion.

\section{Model and method}
\label{Sect:sys_method}

\subsection{System Hamiltonian}

In this paper, we consider negatively charged NV centers in diamond. 
The electronic ground state of the NV center is a triplet state with the spin quantum number $S=1$. 
The NV center spin is coupled to the $^{13}\text{C}$ nuclear spins $\left\{\mathbf I_i\right\}$, which are
spin-$1/2$'s with the natural abundance $p_{\text{nat}}=1.1\%$, randomly distributed on the diamond lattice. 
The full Hamiltonian of the whole system is written as
\begin{equation}
H=H_{\text{NV}}+H_{\text{bath}}+H_{\text{int}}.
\label{Eq:Hamiltonian_Total}
\end{equation}
The Hamiltonians of the NV center and the bath spins in magnetic field $\mathbf{B}$ are 
\begin{eqnarray}
H_{\text{NV}}& = & -\gamma_{\text{e}}\mathbf{B}\cdot\mathbf{S}+\Delta S_{z}^{2},\\
H_{\text{bath}}& = & -\gamma_{\text{n}}\mathbf{B}\cdot\sum_{i}{\mathbf{I}_{i}}+H_{\text{dip}},
\label{Eq:dipolar_H_define}
\end{eqnarray}
where $\gamma_{\text{e}}=-1.76\times 10^{11}\text{~rad~s}^{-1}\text{~T}^{-1}$ and $\gamma_{\text{n}}=6.73\times
10^7\text{~rad~s}^{-1}\text{~T}^{-1}$ are the gyromagnetic ratios of the electron and $^{13}$C nuclear spins,
respectively, $\Delta=2.87$~GHz is the zero-field splitting of the electron spin, and the $z$-axis is taken along the
direction pointing from the nitrogen to the vacancy [see Fig.~\ref{FIG:fig1_schem}(a)] unless specified otherwise. 
In this paper, the magnetic field direction is assumed along the $z$ direction. We focus on the dependence of the
decoherence behavior on the magnetic field strength. 
The effect of the magnetic field direction on the electron spin coherence time has been studied in
Refs.~\onlinecite{Maze2008a,Stanwix2010a,Zhao2011a}. 

In Eq.~(\ref{Eq:dipolar_H_define}), the Hamiltonian $H_{\text{dip}}$ includes the dipolar interactions between nuclear
spins
\begin{eqnarray}
H_{\text{dip}}=\sum_{i<j}{D_{ij}\Big[\mathbf{I}_i \cdot \mathbf{I}_j
-\frac{3\left(\mathbf{I}_i \cdot \mathbf{r}_{ij}\right)\left(\mathbf{r}_{ij} \cdot
\mathbf{I}_j\right)}{r_{ij}^2}\Big]},
\label{Eq:DipInt}
\end{eqnarray}
where $\mathbf{r}_{ij}$ is the displacement from the $i$th nuclear spin
to the $j$th nuclear spin, and the interaction strength is characterized by $D_{ij}=\mu_0\gamma_{\text{n}}^2/(4\pi
r_{ij}^3)$ with $\mu_0$ being the vacuum permeability.

The electron and bath spins are coupled through the hyperfine interaction, which is described by
\begin{eqnarray}
H_{\text{int}}=\mathbf{S}\cdot\sum_{i}{\mathbb{A}_i\cdot\mathbf{I}_i},
\label{hfint}
\end{eqnarray}
where $\mathbb{A}_i$ is the hyperfine interaction tensor for the $i$th nuclear spin, including the isotropic Fermi
contact part and the anisotropic dipolar interaction part.
The Fermi contact part is important for nuclear spins near the electron spin. 
The $^{13}$C nuclear spins appearing in the first few coordinate shells will induce fast ($>\rm{~MHz}$)
electron spin echo envelope modulation (ESEEM).\cite{Childress2006}  
Instead of the ESEEM, this paper focuses on the overall decoherence, which arises from the
coupling to the large number of nuclear spins relatively far away from the NV center. 
For these nuclear spins, the Fermi contact is not important, and the hyperfine interaction takes the dipolar
form
\begin{equation}
\mathbb{A}_i=\frac{\mu_0}{4\pi}\frac{\gamma_{\text{e}}\gamma_{\text{n}}}{r_{i\text{v}}^3}\left(1-\frac{3\mathbf{r}_{i
\text{v}}\mathbf{r}_{i\text{v}}}{r_{i\text{v}}^2}\right),
\end{equation}
where $\mathbf{r}_{i\text{v}}$ is the displacement of the $i$th $^{13}$C spin from the vacancy site.

\subsection{Electron spin coherence}
The total Hamiltonian in Eq.~(\ref{Eq:Hamiltonian_Total}) is rewritten as 
\begin{subequations}
\begin{eqnarray}
H&=&H_{\text{NV}}+\mathbf{b}\cdot\mathbf{S}+H_{\text{bath}},\\
\mathbf{b}&\equiv&\sum_{i}{\mathbb{A}_i\cdot\mathbf{I}_i},
\end{eqnarray}
\label{Eq:Hami_NoiseForm}
\end{subequations}
where the influence of the hyperfine interaction in Eq.~(\ref{hfint}) is expressed in terms of an effective field
$\mathbf{b}$, provided by the nuclear spins, coupling to the NV center spin. 
Either the thermal distribution of the nuclear spin states or the dynamics of nuclear spins causes the fluctuation of
the effective field. 
Thus, the effective field $\mathbf{b}$ is regarded as the noise field, which induces the decoherence of
the NV center spin.

For the magnetic field applied along the $z$ direction, the magnetic quantum number $m$ is a good
quantum number, and the eigenstates of $H_{\text{NV}}$ are denoted as $\vert m \rangle$ with $m=0$, and $\pm 1$, and the corresponding
eigenenergies $\omega_{m}=m^2 \Delta-m \gamma_{\text{e}}B $. With these eigenstates, the hyperfine interaction in Eq.~(\ref{hfint}) is expanded as
\begin{subequations}
\begin{eqnarray}
H_{\text{int}}&=&\sum_{m,n=-1}^{+1}\vert m \rangle \langle n\vert \otimes
b_{m,n},\\
b_{m,n} &=& \mathbf{S}_{m,n} \cdot \mathbf{b}=\sum_i{\mathbf{S}_{m,n} \cdot \mathbb{A}_i\cdot \mathbf{I}_i},
\label{hfint1}
\end{eqnarray}
\end{subequations}
where $\mathbf{S}_{m,n}\equiv\langle m \vert \mathbf{S} \vert n \rangle$. 
The noise operator $b_{m,n}$ contains only the nuclear spin operators. 

For the NV center spin, the diagonal part of the noise operator $b_{m,m}$ induces the energy shift of the level $m$, 
and the off-diagonal part $b_{m,n}$ for $m\neq n$ causes the transition between $\vert m \rangle$ and $\vert n \rangle$. 
Since the NV center has a zero-field splitting in the order of GHz, which is much larger than the typical hyperfine
interaction strength ($< \text{MHz}$), the electron spin can hardly be flipped by the hyperfine interaction.
As a higher-order effect, the hyperfine interaction can still induce indirect coupling between nuclear spins by
virtual flips of the electron spin. 
This indirect coupling can affect the electron spin coherence only when the
nuclear spins involved in the virtual processes are strongly coupled ($>\text{MHz}$) to the electron
spin.\cite{Dutt2007} 
Our previous study\cite{Zhao2011a} has shown that, in most cases that the NV center electron spin
levels are far from the degenerate point, the indirect coupling is much smaller than the intrinsic dipolar interaction
in Eq.~(\ref{Eq:DipInt}), and has negligible effect on the NV center spin decoherence. 
As a consequence, the off-diagonal terms with $m\neq n$ can be safely neglected, and a pure dephasing Hamiltonian is
obtained as
\begin{subequations}
\begin{eqnarray}
H&\approx&H_{\text{NV}} + b_z S_z + H_{\text{bath}}=\sum_{m=-1}^{+1}\vert m \rangle \langle m \vert \otimes H^{(m)},\\
H^{(m)}&\equiv &\omega_{m} + H_{\text{bath}}+b_{m,m},
\end{eqnarray}
\label{Eq:puredephasing}
\end{subequations}
where $H^{(m)}$ is the bath Hamiltonian conditioned on the electron spin state $\vert m\rangle$. 
The operator $b_{m,m}$ is written in the following form
\begin{eqnarray}
b_{m,m}=\sum_i{m \left(\hat{\mathbf{z}} \cdot \mathbb{A}_i\right)\cdot
\mathbf{I}_i} \equiv \sum_i{\mathbf{A}^{(m)}_{i} \cdot \mathbf{I}_i},
\label{Eq:backaction_b}
\end{eqnarray}
where $\mathbf{A}^{(m)}_{i}=m\left(\hat{\mathbf{z}}\cdot
\mathbb{A}_i\right)$ is the effective hyperfine field for the $i$th nuclear spin when
the electron spin is in the state $\vert m \rangle$. 

The hyperfine field in Eq.~(\ref{Eq:backaction_b}) contains two parts, namely, the \textit{isotropic} and
\textit{anisotropic} interactions, according to whether or not the nuclear spin quantum number is conserved. 
The isotropic part contains the coupling along the $z$ direction (i.e. the $A^{(m)}_{i,z} I_{i}^{z}$ term), and provides
frequency shifts to nuclear spins. 
The anisotropic part contains the coupling along the $x$ and $y$ directions (i.e. the $A^{(m)}_{i,x} I_{i}^{x}$ and
$A^{(m)}_{i,y} I_{i}^{y}$ terms), and causes the nuclear spin flipping. 
Since the nuclear spin flipping process does not conserve the Zeeman energy, the effect of the anisotropic coupling on
the decoherence depends on the magnetic field strength. The role of anisotropic coupling under different
magnetic fields will be discussed in detail in Section~\ref{Sect:spin_process_pic} and Section~\ref{Sect:res_disc}.

In the absence of electron spin flipping process in the Hamiltonian Eq.~(\ref{Eq:puredephasing}), the population
of each electron spin states will not change. The coherence of the electron spin at time $t$ is defined by the average
value of the transverse spin component as 
\begin{equation}
L(t)=\frac{\text{Tr}\left[\rho(t) S^{+}\right]}{\text{Tr}\left[\rho(0) S^{+}\right]},
\end{equation}
where $S^{+}\equiv S_{x}+i S_{y}$ and $\rho(t)$ is the density matrix of the total system of electron and bath spins at
time $t$. For the initial time $t=0$, the system is prepared in a product state with $\rho(0)=\rho_{\text{B}}\otimes
|\psi_\text{e}(0)\rangle\langle\psi_\text{e}(0)|$, where $\rho_{\text{B}}$ is the density matrix of bath spins, and the
electron spin is in a superposition state of eigenstates $|m\rangle$ and $|n\rangle$, i.e.
$|\psi_\text{e}(0)\rangle=\alpha |m\rangle+\beta|n\rangle$.

For the temperature much higher than the nuclear spin Zeeman energy ($\lesssim$MHz $\sim$ $\mu$K), the density matrix
$\rho_{\text{B}}$ of the bath spins is well approximated as $\rho_{\text{B}}= \mathbb{I}/2^{M}$, where $M$ is the number
of $^{13}$C spins, and $\mathbb{I}$ is a $2^M\times 2^M$ identity matrix. 
With $N$ DD pulses applied at $t_1, t_2, \ldots, t_N$, which flip the
electron spin, the coherence between $|m\rangle$ and $|n\rangle$ states is expressed
by\cite{Yang2008}
\begin{equation}
L(t)=\text{Tr}_{\text{B}}\left[\cdots e^{-i  H^{(m)}\tau_2}e^{-i  H^{(n)}\tau_1}\rho_{\text{B}}e^{i H^{(m)}\tau_1}e^{i 
H^{(n)}\tau_2} \cdots \right],
\label{Eq:Lgeneral}
\end{equation}
where $\tau_k=t_k-t_{k-1}$ for $t_0=0$ and $t_{N+1}=t$ is the $k$th free evolution interval between the
$(k-1)$th and $k$th pulses [see Fig.~\ref{FIG:fig1_schem}]. 

In this paper, the pulses are assumed to be ideal, which flip the electron spin instantaneously with no error.
We will focus on the periodic DD (PDD) pulse sequences, in which the electron is periodically flipped at times
$t_k=(2k-1)t/2N$. 
This control sequence was shown to be an
efficient way for protecting the electron spin coherence of NV centers.\cite{Lange2010,Ryan2010,Naydenov2011}
The comparison and analysis of the efficiency of various kinds of DD control schemes are beyond the scope of this paper
and will be discussed elsewhere.

\subsection{Cluster-correlation expansion method}

The decoherence of an electron spin in an interacting bath is a many-body problem. 
The CCE method is employed to solve the problem.\cite{Yang2008,Yang2009}

The key idea is that the decoherence function $L(t)$ can be expressed as the product of cluster correlations
\begin{equation}
\label{Eq:CCEgeneral}
{L}\left(t\right)=\prod_{C}\tilde{L}_{C}\left(t\right),
\end{equation}
with the unfactorizable correlation $\tilde{{L}}_{C}\left(t\right)$ of cluster $C$ recursively defined by
\begin{align}
\label{CCEunfactorizable}
\tilde{{L}}_{C}\left(t\right)=\frac{{L}_{C}\left(t\right)}{\prod_{C'\subset C}{\tilde{L}}_{C'}\left(t\right)},
\end{align}
where ${L}_{C}\left(t\right)$ is calculated in the similar way to Eq.~(\ref{Eq:Lgeneral})
\begin{equation}
L_{C}(t)=\text{Tr}_{\text{B}}\left[\cdots e^{-i  H_{C}^{(m)}\tau_2}e^{-i H_{C}^{(n)}\tau_1}\rho_{\text{B}}e^{i
H_{C}^{(m)}\tau_1}e^{i H_{C}^{(n)}\tau_2} \cdots \right],
\label{Eq:CCE_Lc}
\end{equation}
where only the interaction within the cluster $C$ is included in the bath conditional Hamiltonian $H^{(m)}_{C}$.
In realistic calculation, the expansion is truncated at a certain size $K$ of clusters (defined as the number of spins
in the cluster), i.e. $|C|\le K$, and the result is denoted as CCE-$K$. 
The final electron spin coherence is obtained by increasing the value of $K$ until the results get converged. 
The CCE convergence order depends on the underlying microscopic decoherence mechanisms in the timescale of interest,
Thus, CCE is not only an efficient calculation method, but also provides a tool for analyzing the dominating
decoherence mechanisms. 
Typically, the truncation up to four-spin clusters, i.e. CCE-4, can give a convergent result. For the electron spin
coherence under many-pulse DD controls, as the coherence time is prolonged to $\sim 10$~ms, the calculations are convergent at CCE-6.

\section{Physical processes}
\label{Sect:spin_process_pic}

For NV centers in nuclear spin baths, the noises can be of either thermal or quantum nature.\cite{Zhao2011b,Huang2011arXiv} 

At room temperature, the random orientations of the nuclear spins result in thermal noises. 
The density matrix of nuclear spin is written as 
\begin{equation}
\rho_{\text{B}}=\frac{1}{2^M}\sum_{J} |J\rangle\langle J|,
\end{equation}
where $|J\rangle$ is chosen as the eigenstate of the noise operator $b_z$ [see Eqs.~(\ref{Eq:Hami_NoiseForm}) and
(\ref{Eq:puredephasing})] with
\begin{equation}
b_z |J\rangle = b_J|J\rangle.
\end{equation}
The thermal distribution of the nuclear spin states $|J\rangle$ leads to the random Overhauser field $b_J$ for the NV
center spin, which is the thermal noise.
The thermal noises are so strong that they induce the NV center decoherence in several
$\mu$s. However, the thermal noises are quasi-static and can be completely eliminated by spin echo.
Then, the dynamical noises caused by the quantum evolution of the nuclear spins become important for NV centers under
DD controls.

The dynamical noises arise from the quantum evolution of the nuclear spins.
Since the noise operator $b_z$ does not commute with the bath Hamiltonian $H_{\text{bath}}$, i.e. $[b_z,
H_{\text{bath}}]\neq 0$, the eigenstates $|J\rangle$ of the noise field are in general not the eigenstates of the total
Hamiltonian.
During the evolution of the system, the nuclear spins initially in state $|J\rangle$ will be brought to the
superposition of eigenstates of $b_z$ corresponding to different Overhauser field $b_J$.
Thus, the evolution of the nuclear spins causes the dynamical quantum noises of the NV center spin levels. 
In order to investigate the decoherence mechanisms under DD controls, it is critical to understand the elementary
processes in the nuclear spin evolution. Since the hyperfine interaction and the nuclear spin dipolar interaction
involve one or two nuclear spins, respectively, the basic physical processes occurring in the nuclear spin bath are the single nuclear spin precession and
the nuclear spin pair flip-flops.

\subsection{Single nuclear spin dynamics}
\label{SubSect:SingleSpin}

\begin{figure}[tb]
  \includegraphics[width= \columnwidth]{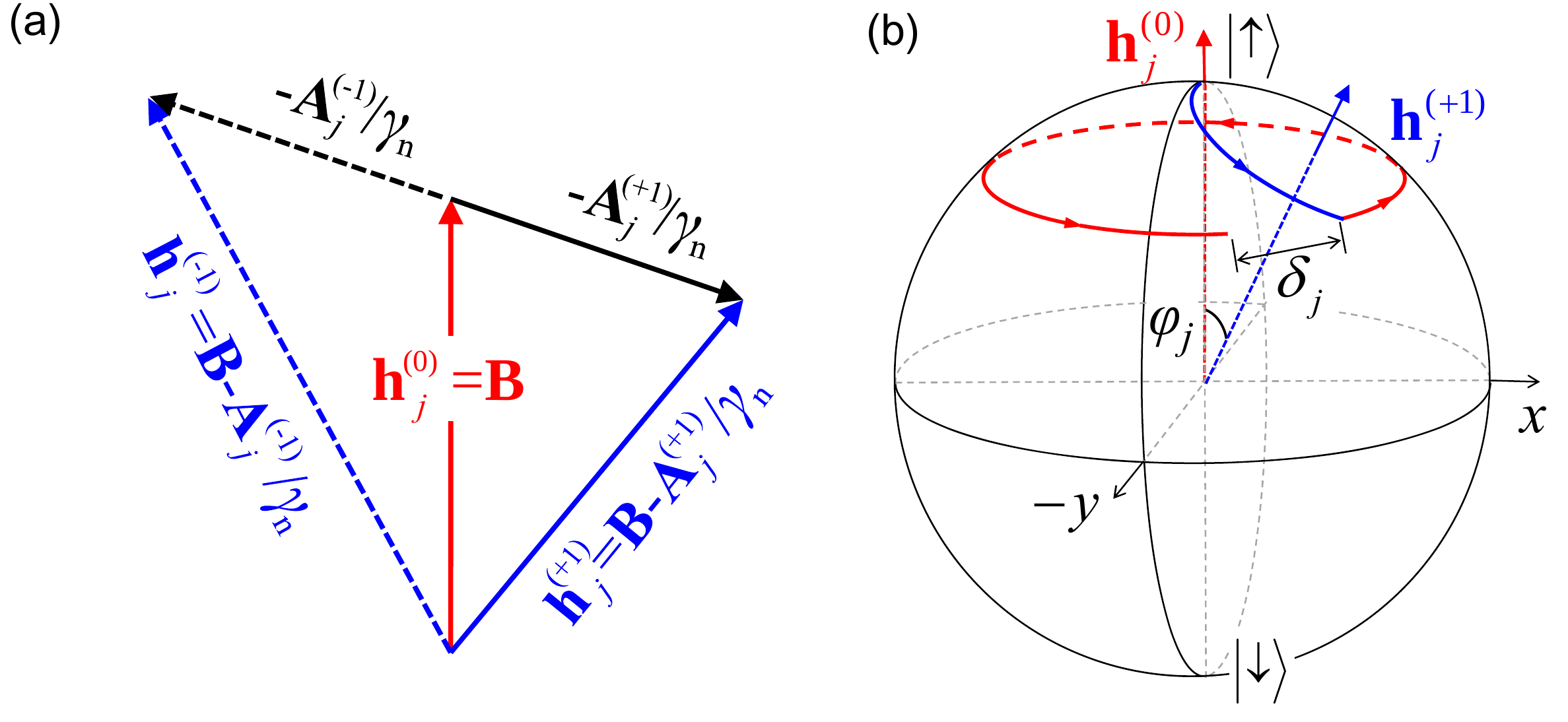}
  \caption{(a) Effective fields $\mathbf{h}_{j}^{(m)}$ for the $j$th single nuclear spin in a magnetic field $\mathbf{B}$
  and a hyperfine field $\mathbf{A}_j^{(m)}$ conditioned on the electron spin state $|m\rangle$. 
  (b) Bloch sphere representation of the bifurcated evolution of the $j$th nuclear spin under a single pulse control
  (Hahn echo). The single nuclear spin is driven by the effective fields $\mathbf{h}_j^{(0)}$ and $\mathbf{h}_j^{(+1)}$.
  The distance $\delta_j$ between the trajectories measures the decoherence  due to the $j$th nuclear spin.}
\label{FIG:singlespin}
\end{figure}

In this subsection, we investigate the quantum evolution of the single-spin clusters.
The conditional Hamiltonian of the $j$th nuclear spin is
\begin{subequations}
\label{Eq:SingleSpinHami}
\begin{eqnarray}
H_{j}^{(m)}&=&-\gamma_{\text{n}}\mathbf{h}_j^{(m)}\cdot\mathbf{I}_j,\\
\mathbf{h}_j^{(m)}&=&\mathbf{B}-\mathbf{A}_j^{(m)}/\gamma_{\text{n}}.
\end{eqnarray}
\end{subequations}
The geometric picture of the effective field $\mathbf{h}_j^{(m)}$ is shown in Fig.~\ref{FIG:singlespin}(a), 
and the Bloch sphere representation of the single spin precession is illustrated in
Fig.~\ref{FIG:singlespin}(b). 
The north and south poles represent the basis states $|\uparrow\rangle$ and $|\downarrow\rangle$, respectively.  For the
electron spin in state $|0\rangle$, the hyperfine field vanishes, and the nuclear spin rotates about the applied
magnetic field, i.e. $\mathbf{h}_j^{(0)}= \mathbf{B}$. For the electron spin state $|\pm 1\rangle$, the effective field
is $\mathbf{h}_j^{(\pm 1)}= \mathbf{B}- \mathbf{A}_j^{(\pm 1)}/\gamma_{\text{n}}$. 

Because of the anisotropic nature of the hyperfine interaction, the effective
fields $\mathbf{h}_j^{(0)}$ and $\mathbf{h}_j^{(\pm 1)}$ are in general not parallel to each other. 
This is different from the cases of the QDs and shallow donors. In those systems, the hyperfine interaction is
dominated by the isotropic Fermi contact part, so the diagonal hyperfine field is always in the same direction as the applied
magnetic field. As a result, in QDs and shallow donors, the single nuclear spin dynamics does not significantly
contribute to the electron spin decoherence. 

In the NV center system, the precession of the $j$th nuclear spin about the
unparalleled effective fields $\mathbf{h}_j^{(m)}$ and $\mathbf{h}_j^{(m')}$ (with $m,m'=0$ or $\pm 1$, and $m\neq m'$)
gives the bifurcated paths of the nuclear spins, and the electron spin coherence $L_{j}(t)$ is determined by the distance
\begin{equation}
\delta_j\equiv \sqrt{1-|\langle\psi^{(m)}(t)|\psi^{(m')}(t)\rangle|^2}
\end{equation}
between the trajectories on the Bloch
sphere,\cite{Yao2007,Liu2007} where $|\psi^{(m)}(t)\rangle$ and $|\psi^{(m')}(t)\rangle$ are the nuclear spin state
driven by the effective field $\mathbf{h}_j^{(m)}$ and $\mathbf{h}_j^{(m')}$, respectively.

According to Eq.~(\ref{Eq:CCE_Lc}), the $j$th single nuclear spin contribution to the FID and Hahn echo of the
coherence between $|0\rangle$ and $|+1\rangle$ states is calculated as\cite{Maze2008a}
\begin{subequations}
\begin{align}
&{L}_{j,\text{FID}}\left(t\right)=\cos\frac{\theta_j^{(0)}}{2}\cos\frac{\theta_j^{(+1)}}{2}+\sin\frac{\theta_j^{(0)}}{2}\sin\frac{\theta_j^{(+1)}}{2}\cos\varphi_j,\\
&{L}_{j,\text{Hahn}}\left(t=2\tau\right)=1-2\sin^2\varphi_j\sin^2\frac{\theta_j^{(0)}}{2}\sin^2\frac{\theta_j^{(+1)}}{2},
\end{align}
\label{Eq:SingleSpinCoherenceHahnEcho}
\end{subequations}
where $\theta_j^{(m)}=\gamma_{\text{n}}h_j^{(m)}\tau$ and $\varphi_j$ is the angle between $\mathbf{h}_j^{(0)}$ and
$\mathbf{h}_j^{(+1)}$ (see Fig.~\ref{FIG:singlespin}). Under $N$-pulse PDD (PDD-$N$) controls, the decoherence at $t=2N\tau$
due to $j$th nuclear spin is expressed as
\begin{equation}
L_{j,N}(t)=
\begin{dcases}
1-2\sin^2\alpha\sin^2\left(\frac{k\theta}{2}\right), & N=2k\\
L_{j,\text{Hahn}}(2\tau)L_{j,2k}(4k\tau)+L_{\text{corr}}(\tau), & N=2k+1
\label{Eq:SingleSpinCoherencePDD}
\end{dcases},
\end{equation}
for $k=1,2,\ldots$. The angles $\alpha$ and $\theta$ are defined as
\begin{subequations}
\label{Eq:PDDangles}
\begin{eqnarray}
\label{Eq:PDDanglesA}
&&\tan\alpha=\frac{\sin\frac{\theta_j^{(0)}}{2}\sin\frac{\theta_j^{(+1)}}{2}\sin{\varphi_j}}
{\cos\frac{\theta_j^{(0)}}{2}\cos\frac{\theta_j^{(+1)}}{2}-\sin\frac{\theta_j^{(0)}}{2}\sin\frac{\theta_j^{(+1)}}{2}\cos\varphi_j},\\
\label{Eq:PDDanglesB}
&&\cos\frac{\theta}{2}=\cos\theta_j^{(0)}\cos\theta_j^{(+1)}-\sin\theta_j^{(0)}\sin\theta_j^{(+1)}\cos\varphi_j.
\end{eqnarray}
\end{subequations}
In Eq.~(\ref{Eq:SingleSpinCoherencePDD}), the coherence under PDD-$(2k+1)$ control is written as the product of
Hahn echo and PDD-$2k$ control with an additional correction term $L_{\text{corr}}(\tau)$. 
The expression of $L_{\text{corr}}(\tau)$, the derivation of
Eqs.~(\ref{Eq:SingleSpinCoherenceHahnEcho})-(\ref{Eq:PDDangles}) and the geometric pictures can be found in Appendix.

The single nuclear spin contribution to the electron spin decoherence depends on the magnitude of the applied magnetic
field. 
In the absence of magnetic field, the effective fields are $\mathbf{h}_j^{(0)}=0$ and
$\mathbf{h}_j^{(+1)}=-\mathbf{A}_j^{(+1)}/\gamma_{\text{n}}$, and all the evolution operators in Eq.~(\ref{Eq:CCE_Lc})
mutually commute. Therefore the single spin dynamics does not contribute to the electron spin decoherence at the
echo time. For weak magnetic field ($B\lesssim 1$~Gauss), the precession angle $\theta_j^{(0)}$ about the magnetic field is small
within the relevant decoherence timescale ($\tau \ll \text{ms}$). In this case, the single spin dynamics of each nuclear
spin gives a minor contribution to the electron spin decoherence, which is characterized by the distinguishability
$\delta_j^2$ between the two trajectories, i.e. $\delta_j^2 \propto \gamma_{\text{n}}^2B^2\tau^2\ll 1$.

In the opposite limit that the nuclear spin Zeeman energy much is larger than the hyperfine interaction, the single
nuclear spin contribution to the decoherence is suppressed. 
According to the Bloch sphere picture shown in Fig.~\ref{FIG:singlespin}(b) and
Eqs.~(\ref{Eq:SingleSpinCoherenceHahnEcho})-(\ref{Eq:PDDangles}), the single spin contribution to the decoherence
is controlled by the angle $\varphi_j$ between the two effective fields. Increasing the magnetic field so that $\gamma_{\text{n}}B\gg A_j^{(+1)}$, the pseudo-fields are approximately parallel with each other, i.e.
$\sin\varphi_j \ll 1$. In this case, the two bifurcated evolution paths do not have significant distinguishability. In
other words, the decoherence due to single nuclear spin dynamics is small in the strong field limit.

For a medium magnetic field, the single nuclear spin precession induces the oscillations of the electron spin
coherence. As shown in Eqs.~(\ref{Eq:SingleSpinCoherenceHahnEcho})-(\ref{Eq:PDDangles}), there are
two recovery periods of the total evolution time $T_{j}^{(m)}=4N\pi/(\gamma_{\text{n}}h_j^{(m)})$ for $m=0$ and $+1$.
These two recovery periods correspond to the nuclear spin precessing about the effective fields $\mathbf{h}_j^{(m)}$ for full circles. 
Since the effective field $\mathbf{h}_j^{(0)}=\mathbf{B}$ is actually the applied magnetic field, the recover period
$T_{j}^{(0)}=4N\pi / (\gamma_{\text{n}}B)$ is the same for all the nuclear spins. 
The simultaneous recovery of coherence for all nuclear spins results in the periodical coherence
revival effect, which has been experimentally observed in the Hahn echo
case.\cite{VanOort1990,Childress2006,Stanwix2010a}

\begin{figure}[tb]
  \includegraphics[width= \columnwidth]{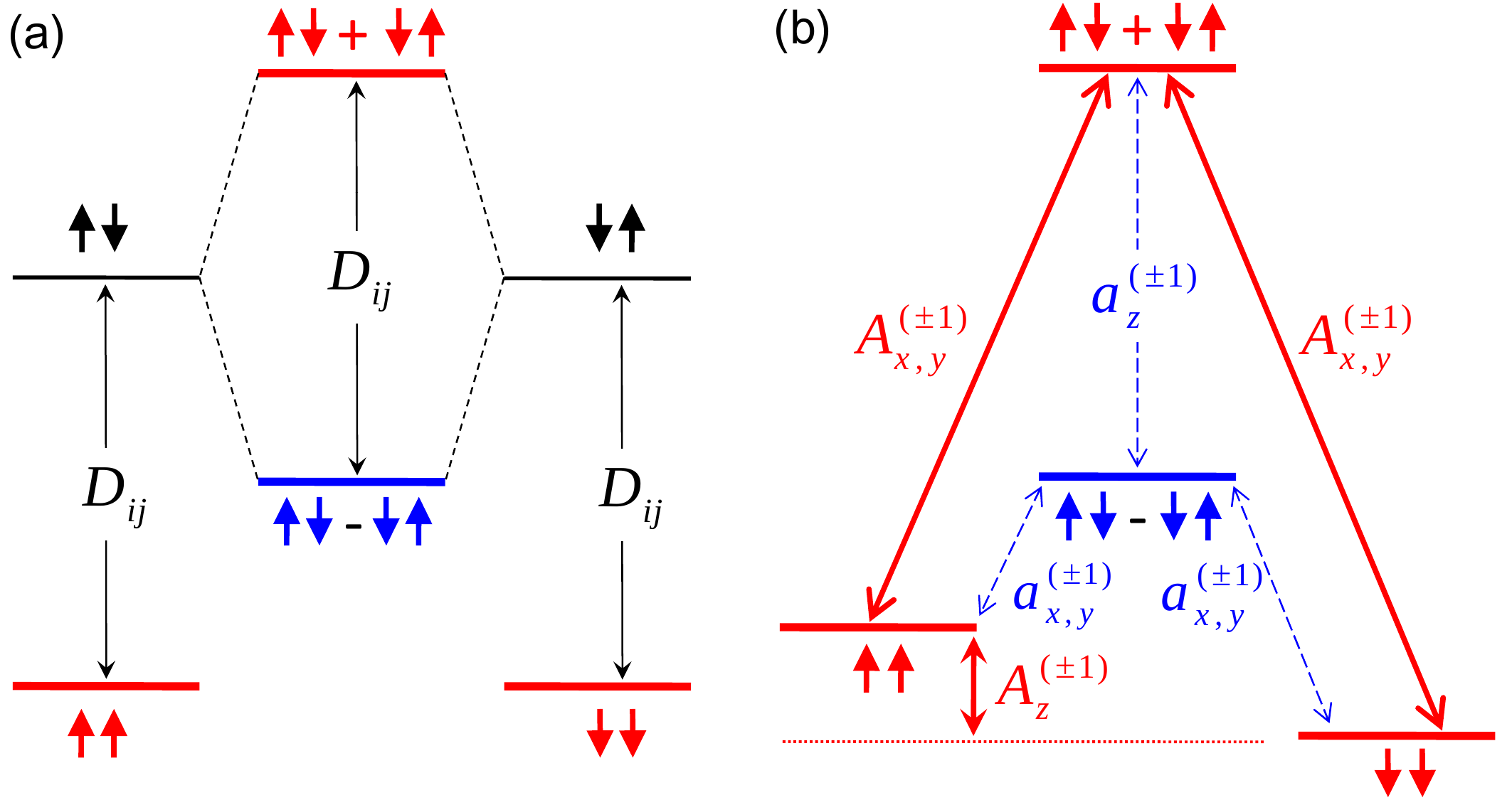}
  \caption{(a) Energy levels and eigenstates of two-nuclear spin Hamiltonian $H_{\{i,j\}}^{(0)}$ under zero magnetic field for
  the NV center spin in state $|0\rangle$.
  (b) The energy shift and the transitions induced by the hyperfine fields. The transitions induced by the average
  hyperfine field $\mathbf{A}^{(\pm1)}$ (the red thick arrows) are much stronger than those induced by the hyperfine
  field difference $\mathbf{a}^{(\pm1)}$ (the blue dashed arrows). 
  }
  \label{FIG:fig3_nonSecular}
\end{figure}

\subsection{Nuclear spin pair dynamics}
\label{Sect:SpinPair}
In a two-spin cluster, the two nuclear spins are correlated due to the dipolar interaction.
In general, there are two types of processes driven by the dipolar interaction, namely, 
(i) the non-secular spin flipping [e.g., the processes described by terms like $I_{i}^{z}I_{j}^{+}$ and
$I_{i}^{+}I_{j}^{+}$], and (ii) the secular spin flip-flop [e.g., the process described by $I_{i}^{+}I_{j}^{-}$]. 
The non-secular flipping does not conserve the nuclear Zeeman energy while the secular flip-flop does.
Consequently, they have different dynamic behaviors under different magnetic fields.

\subsubsection{Non-secular spin flipping}
\label{subsubsect:nonsecular}

\begin{figure}[tb]
  \includegraphics[width= \columnwidth]{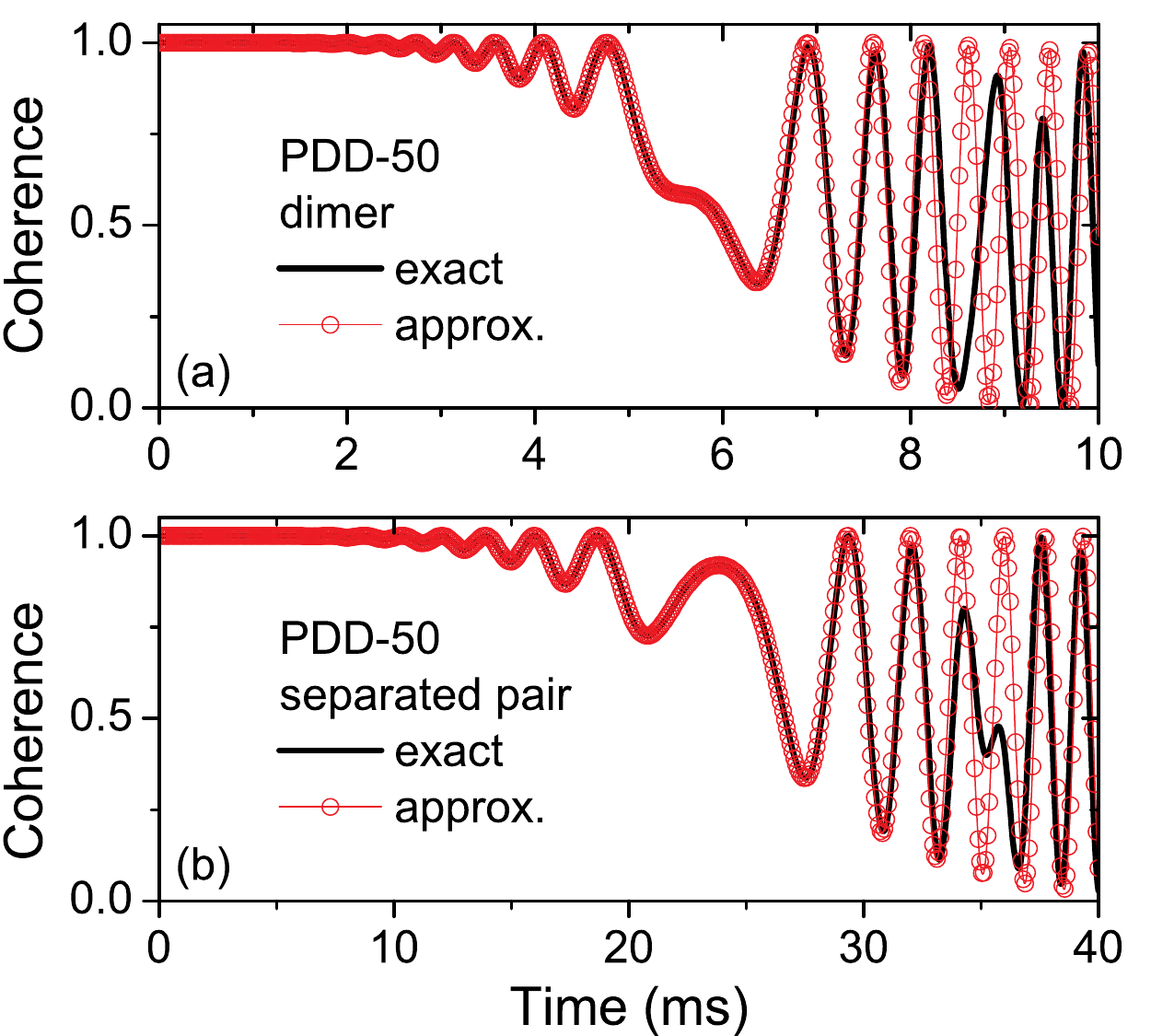}
  \caption{(a) The electron spin decoherence between $|+1\rangle$ and $|-1\rangle$ states induced by a nuclear spin dimer under
  PDD-50 control and zero magnetic field. The black line is the exact calculation result using Eq.~(\ref{Eq:CCE_Lc}), and the red line with symbols is
  obtained by the pseudo-spin model [Eq.(\ref{Eq:SpinJ})]. (b) The same as (a), but for the two nuclear spins separated
  by 2.52~$\text{\AA}$ (the next nearest neighboring pair). Notice the different timescales of
  the horizontal axes. The separated pair has much slower dynamics than the dimer.}
\label{FIG:fig3_nonSecular_data}
\end{figure}

The non-secular process occurs when the nuclear Zeeman energy cost is less than or comparable to the dipolar interaction
strength, i.e. $\gamma_{\text{n}}B\lesssim D_{ij}$. Since the dipolar interactions between $^{13}$C nuclear spins are at
most $\sim \rm {kHz}$ (for the closely bonded pairs), the non-secular process will be activated in the weak magnetic
field $\lesssim 1$~Gauss.

For the sake of clarity, we consider first the zero magnetic field case.
For the electron spin in the $|0\rangle$ state, the Hamiltonian of the
nuclear spin pair only contains the dipolar interaction. 
By choosing the quantization axis along the direction of relative displacement of the two spins,
i.e. $\hat{\mathbf{z}}\parallel \mathbf{r}_{ij}$, the Hamiltonian $H_{\{i,j\}}^{(0)}$ is written as
\begin{equation}
H_{\{i,j\}}^{(0)}=\frac{D_{ij}}{2}\left[\left(
I_{i}^{+}I_{j}^{-}+I_{i}^{-}I_{j}^{+}\right)-4I_{i}^{z}I_{j}^{z}\right],
\label{Eq:pairCondHami0}
\end{equation}
where $I_{i}^{\pm}=I_{i}^{x}\pm i I_{i}^{y}$. 
Since the Hamiltonian $H_{\{i,j\}}^{(0)}$ commutes with the total spin angular momentum
$(\mathbf{I}_i+\mathbf{I}_j)^2$ of the two $^{13}$C nuclear spin-$1/2$'s, the eigenstates of $H_{\{i,j\}}^{(0)}$ are
classified as the triplet states, $|T_{+1}\rangle=|\uparrow\uparrow\rangle$,
$|T_{0}\rangle=\left(|\uparrow\downarrow\rangle+|\uparrow\downarrow\rangle\right)/\sqrt{2}$ and
$|T_{+1}\rangle=|\downarrow\downarrow\rangle$, and singlet state
$|S\rangle=\left(|\uparrow\downarrow\rangle-|\uparrow\downarrow\rangle\right)/\sqrt{2}$, according to their total
angular momentum. The eigenstates and the corresponding energy levels are shown in
Fig.~\ref{FIG:fig3_nonSecular}(a).

\begin{figure}[tb]
  \includegraphics[width= \columnwidth]{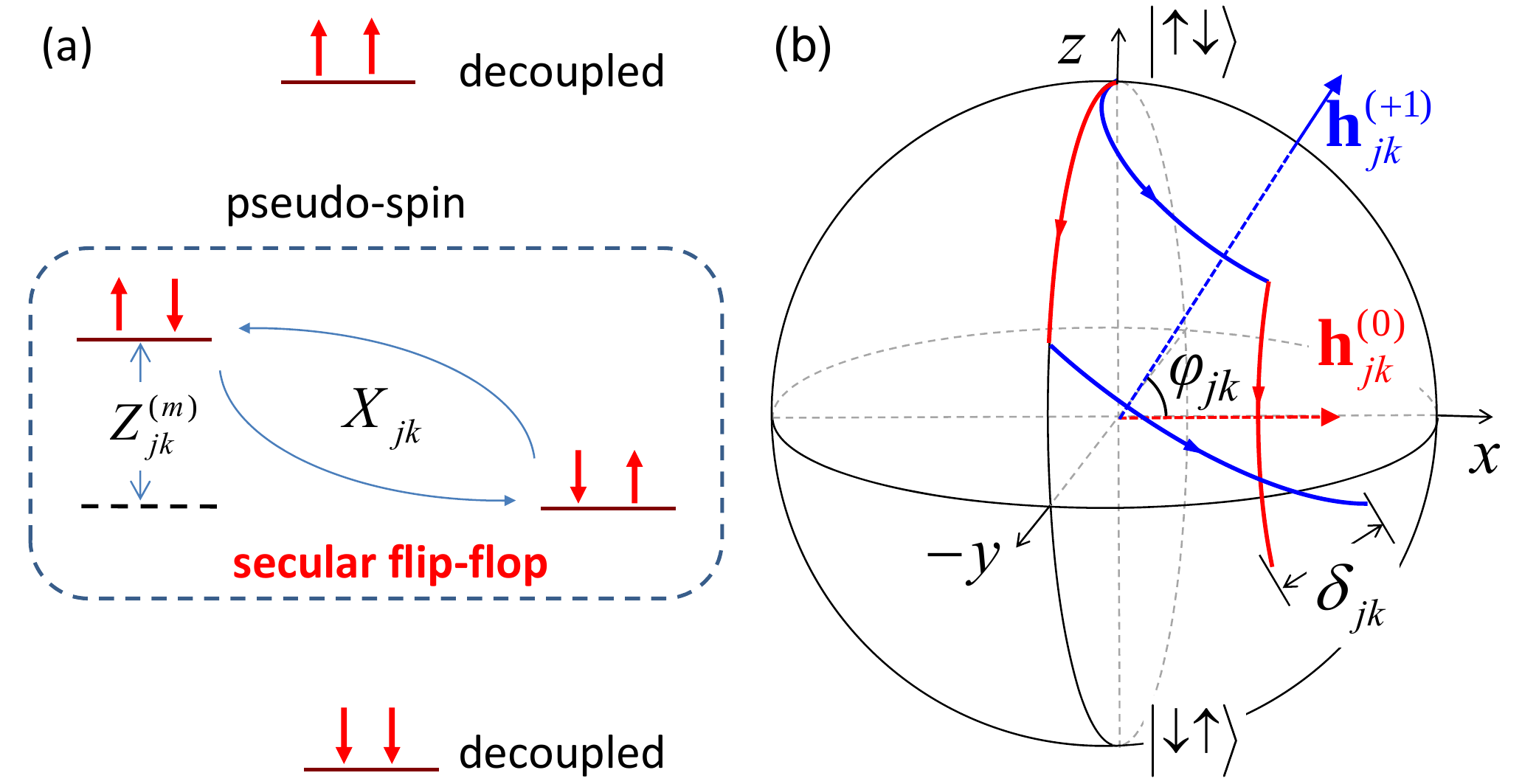}
  \caption{(a) Energy levels of two-spin cluster under a strong magnetic field. The two polarized states are decoupled and
  the dynamics of the two unpolarized states is described by a pseudo-spin. 
  (b) The bifurcated evolution paths of the pseudo-spin on the Bloch sphere. The pseudo-spin is driven by
  the pseudo-fields $\mathbf{h}_{jk}^{(0)}$ and $\mathbf{h}_{jk}^{(+1)}$. The distance $\delta_{jk}$ between the
  trajectories measures the decoherence  due to the nuclear spin pair.}
  \label{FIG:fig4_BlochSphere}
\end{figure}

For the electron spin in the $|\pm 1\rangle$ states, besides the dipolar interaction between nuclear spins, the
hyperfine interactions provide effective fields for the nuclear spins. The conditional Hamiltonian reads
\begin{subequations}
\begin{eqnarray}
H_{\{i,j\}}^{(\pm 1)}&=&H_{\{i,j\}}^{(0)}+\mathbf{A}_i^{(\pm1)}\cdot \mathbf{I}_i+\mathbf{A}_j^{(\pm1)}\cdot
\mathbf{I}_j\\
&\equiv&H_{\{i,j\}}^{(0)}+\mathbf{A}_{\{i,j\}}^{(\pm1)}\cdot
\left(\mathbf{I}_i+\mathbf{I}_j\right)+\frac{\mathbf{a}_{\{i,j\}}^{(\pm1)}}{2}\cdot \left(\mathbf{I}_i-\mathbf{I}_j\right).
\label{Eq:pairCondHami1}
\end{eqnarray}
\end{subequations}
In Eq.~(\ref{Eq:pairCondHami1}), we have expressed the hyperfine interaction in terms of the average
hyperfine field $\mathbf{A}_{\{i,j\}}^{(m)}\equiv\left(\mathbf{A}^{(m)}_i+\mathbf{A}^{(m)}_j\right)/2$, and the
hyperfine field difference $\mathbf{a}_{\{i,j\}}^{(m)}\equiv\mathbf{A}^{(m)}_i-\mathbf{A}^{(m)}_j$.

In general, due to the presence of the hyperfine fields, the states $|S\rangle$ and
$|T_{0,\pm1}\rangle$ are not the eigenstates of $H_{\{i,j\}}^{(\pm1)}$.
Because the coupling to the average hyperfine field $\mathbf{A}_{\{i,j\}}^{(\pm1)}$ conserves the total angular
momentum, $\mathbf{A}_{\{i,j\}}^{(\pm1)}$ will induce the transitions and the level shift within the triplet subspace. 
The coupling to the hyperfine difference $\mathbf{a}_{\{i,j\}}^{(\pm1)}$ breaks the conservation of total angular
momentum, and causes the transitions between the singlet and triplet states [see Fig.~\ref{FIG:fig3_nonSecular}(b)]. 
In particular, the transition between $|S\rangle$ and $|T_{0}\rangle$ states, which conserves the $z$ component of the
total angular momentum, is due to the secular spin flip-flop.

For relatively near-neighboring nuclear spin pairs (with the inter-nucleus distance of several $\text{\AA}$), the
average hyperfine field is usually much larger than the hyperfine field difference, i.e.
$|\mathbf{A}_{\{i,j\}}^{(\pm1)}|\gg|\mathbf{a}_{\{i,j\}}^{(\pm1)}|$. With this observation, the singlet state
$|S\rangle$ is effectively decoupled from the triplet states $|T_{0,\pm}\rangle$ within the relevant timescale. 
The triplet states $|T_{0,\pm}\rangle$ form a three-level system, whose dynamics is described by the motion of a
pseudo-spin-1 $\mathbf{J}_{\{i,j\}}$. 
The conditional Hamiltonians in Eqs.~(\ref{Eq:pairCondHami0}) and (\ref{Eq:pairCondHami1}) are expressed in terms of the
pseudo-spin $\mathbf{J}_{\{i,j\}}$ as
\begin{eqnarray}
H_{\{i,j\}}^{(m)}\approx-\frac{3D_{ij}}{2}\left(J_{\{i,j\}}^{z}\right)^2+\mathbf{A}_{\{i,j\}}^{(m)}\cdot
\mathbf{J}_{\{i,j\}}.
\label{Eq:PseudoSpinJ}
\end{eqnarray}
The pseudo-spin model described in Eq.~(\ref{Eq:PseudoSpinJ}) is obtained in the zero magnetic field. 
The finite homogeneous magnetic field does not cause the mixing between singlet and triplet states, since it conserves
the total spin. For the weak but non-zero magnetic field $0<\gamma_{\text{n}}B\lesssim D_{ij}$, the pseudo-spin model
for the non-secular process is still valid but the average hyperfine field $\mathbf{A}_{\{i,j\}}^{(m)}$ should be replaced by
$\mathbf{A}_{\{i,j\}}^{(m)}-\gamma_{\text{n}}\mathbf{B}$ in Eq.~(\ref{Eq:PseudoSpinJ}).

With this pseudo-spin model, the electron spin decoherence (e.g., between $|+1\rangle$ and $|-1\rangle$ electron
spin states) induced by a nuclear spin pair under DD controls is expressed as
\begin{equation}
L_{\{i,j\}}(t)=\frac{1}{4}+\frac{1}{4}\text{Tr}\left[\cdots e^{-i  H_{\{i,j\}}^{(+1)}\tau_2}e^{-i  H_{\{i,j\}}^{(-1)}\tau_1}e^{i
H_{\{i,j\}}^{(+1)}\tau_1}e^{i H_{\{i,j\}}^{(-1)}\tau_2} \cdots \right].
\label{Eq:SpinJ}
\end{equation} 
As an example, the nuclear spin pair induced decoherence under a 50-pulse PDD (PDD-50) control is shown in
Fig.~\ref{FIG:fig3_nonSecular_data}. 
Within the electron spin decoherence time [$t\lesssim 6$~ms under PDD-50 control, see the full calculation results for
the decoherence time in Fig.~\ref{FIG:fig9_DDzeroB}], the pseudo-spin model reproduces very well the exact
result. The deviation from the exact result for larger time arises from the neglected
dynamics driven by the hyperfine field difference $\mathbf{a}_{\{i,j\}}^{(m)}$. 
In particular, this pseudo-spin model shows that, in the weak field regime $\gamma_{\text{n}}B<1$~Gauss, 
the secular spin flip-flop processes induced by the $z$ component of $\mathbf{a}_{\{i,j\}}^{(m)}$ have negligible effect
on the spin pair dynamics within the relevant timescale.

Furthermore, the decoherence is sensitive to the dipolar interaction strength $D_{ij}$. 
The dynamics of a separated nuclear spin pair [see Fig.~\ref{FIG:fig3_nonSecular_data}(b)] is much slower than that of
a dimer since the former has much weaker dipolar interaction.
In Section~\ref{SubSubSect:WeakField}, we will show that the separated pairs have small contribution to the electron
spin decoherence in the weak field regime.

Thus, we conclude that, although both secular and non-secular processes can occur under weak magnetic field, the
non-secular nuclear spin flipping is the dominant mechanism of electron spin decoherence,
since it involves the average hyperfine field of the two nuclear spins, while the secular process involves the much weaker hyperfine field
difference.

\subsubsection{Secular spin flip-flop}
\label{SubSubSect:Secular}

When the Zeeman splitting is much larger than the dipolar interaction strength ($\gamma_{\text{n}}B\gg D_{ij}$) and the
hyperfine coupling ($\gamma_{\text{n}}B\gg A_i^{(m)}$ and $A_j^{(m)}$), the energy non-conserving process is suppressed
by the Zeeman energy cost. The polarized states $\left|\uparrow\uparrow\right\rangle $ and
$\left|\downarrow\downarrow\right\rangle$ are decoupled during the quantum evolution, and don't contribute to
the electron spin decoherence (except for the static inhomogeneous broadening effect).
The unpolarized states $\left|\uparrow\downarrow\right\rangle $ and
$\left|\downarrow\uparrow\right\rangle$ form a two-dimensional invariant subspace [see
Fig.~\ref{FIG:fig4_BlochSphere}(a)]. The nuclear spin pair in this subspace is mapped to a pseudo-spin-$1/2$, i.e., $\left|\uparrow\downarrow\right\rangle
\mapsto\left|\Uparrow\right\rangle $ and $\left|\downarrow\uparrow\right\rangle \mapsto\left|\Downarrow\right\rangle $.
Similar to the single nuclear spin case, the coherent evolution of the pseudo-spin is understood using the 
Bloch sphere picture shown in Fig.~\ref{FIG:fig4_BlochSphere}(b).

The nuclear spin flip-flop driven by the dipolar interaction gives rise to the
transition between the pseudo-spin states $\left|\Uparrow\right\rangle $ and $\left|\Downarrow\right\rangle $.
The transition rate is calculated through the nuclear spin dipolar interaction as
\begin{equation}
X_{ij}\equiv \langle
\uparrow\downarrow|H_{\text{dip}}|\downarrow\uparrow\rangle=\frac{1}{2}D_{ij}\left(1-3\cos^2\theta_{ij}\right),
\end{equation}
where $\theta_{ij}$ is the angle between the pair orientation and the external magnetic field.

The difference between the hyperfine fields projected along the direction of the external magnetic field induces an energy
cost of the flip-flop
\begin{equation}
Z_{ij}^{(m)}\equiv \langle
\uparrow\downarrow|b_{m,m}|\uparrow\downarrow\rangle=\hat{\mathbf{z}}\cdot(\mathbf{A}_{i}^{(m)}-\mathbf{A}_{j}^{(m)}).
\end{equation}
Thus the flip-flop is mapped to the precession of a pseudo-spin $\boldsymbol{\sigma}$
about a pseudo-field $\mathbf{h}_{ij}^{(m)}=\left(X_{ij},0,Z^{(m)}_{ij}\right)$
conditioned on the electrons spin state $\vert m\rangle$.

The conditional Hamiltonian of the pseudo-spin reads
\begin{equation}
H^{(m)}_{\sigma}=\frac{1}{2}\mathbf{h}_{ij}^{(m)}\cdot\boldsymbol{\sigma}
=\frac{1}{2}\left(X_{ij}\sigma_{x}+Z_{ij}^{\left(m\right)}\sigma_{z}\right).
\end{equation}
A typical evolution path of the pseudo-spin is shown in Fig.~\ref{FIG:fig4_BlochSphere}(b). 

Now that the pseudo-spin picture for the secular flip-flop of the nuclear spin pair is quite similar to the picture of
single nuclear spin precession, the pair spin dynamics under DD controls can be understood in the similar way.  
The decoherence contributed by the nuclear spin pair in large magnetic
field is obtained in the same form as in Eq.~(\ref{Eq:SingleSpinCoherenceHahnEcho}) with the effective fields
$\mathbf{h}_j^{(m)}$ replaced by the pseudo-fields $\mathbf{h}_{ij}^{(m)}$, and the angle $\varphi_j$ replaced by
the angle between the pseudo-fields $\varphi_{ij}=\arctan(Z_{ij}/X_{ij})$.

In most cases, for the inter-nucleus distance of several angstroms, the
flip-flop transition rate $X_{ij}$ ($\lesssim 100$~Hz) is much less
than the energy cost $Z_{ij}^{(+1)}$ ($\sim $~kHz) due to the hyperfine field
difference. 
In this case, the pair contributes only a small fraction to the electron spin decoherence in the short time limit
$\tau\ll X_{ij}^{-1}$. 
When the time gradually increases and the short time limit is not well satisfied, such kind of separated pairs will
become important.
A large number of such pairs give rise to a relatively smooth decay profile of the electron spin coherence. 
In Section~\ref{SubsubSect:StrongMag}, we will show that such kind of pair flip-flop process is the dominant decoherence mechanism
in the strong magnetic field regime.

In the opposite limit, i.e. $Z_{ij}^{(m)}\ll X_{ij}$, which corresponds to dimers
located far away (e.g., several nanometers) from the NV center, two pseudo-fields $\mathbf{h}_{ij}^{(0)}$ and
$\mathbf{h}_{ij}^{(+1)}$ are approximately parallel, i.e., the angle $\varphi_{ij}\approx 0$. 
Such kind of remote dimers give small contribution to the electron spin coherence, and are not important
under the strong magnetic field. 

Between these two limiting cases, a dimer not far (e.g.,$1.5$~nm) from the NV center can
have dipolar interaction comparable to the hyperfine field difference, i.e. $X_{ij}\sim Z_{ij}^{(+1)}\sim \text{kHz}$. 
Such dimers will induce \textit{coherent} oscillations on the relatively smooth decoherence background. 
The application of the strong oscillation features in the atomic-scale magnetometry is discussed in
Ref.~\onlinecite{Zhao2011a}.

\section{Results and discussions}
\label{Sect:res_disc}

This section shows the calculation results of the NV center electron spin coherence. 
The NV center spin coherence has different decoherence behavior in different magnetic fields and at different
timescales. The various decoherence behavior arises from the elementary processes discussed in the previous Section. 
To identify the role of each elementary process in different magnetic fields, 
the following methods are used in analyzing the NV center spin coherence. 

First, in order to find out what kind of clusters are responsible to the electron spin decoherence, 
with the help of CCE, we resolve the contribution to the decoherence of nuclear spin clusters of different sizes. 
The CCE-1 will give the main contribution if the decoherence is caused mainly by the single spin dynamics described in
Section~\ref{SubSect:SingleSpin}. While if the $n$-spin correlation is the dominant decoherence mechanism, the CCE does
not give the convergent result until the clusters of size $n$ are included. 
Second, the role of different types of interaction, including the isotropic and anisotropic terms of the hyperfine
coupling [see Eq.~(\ref{Eq:backaction_b})], and the secular and non-secular terms of the nuclear spin dipolar interaction, are
studied. To this end, we compare the exact results, which are obtained by solving the full Hamiltonian (\ref{Eq:Hamiltonian_Total}), with the
results calculated by applying the following approximations: 
(i) the \textit{isotropic approximation}, in which the anisotropic terms in hyperfine coupling such as $S_zI_j^{x}$ and
$S_zI_j^{y}$ are neglected, and (ii) the \textit{secular approximation}, in which the non-secular terms in nuclear spin
dipolar interaction such as $I_i^{z}I_j^{+}$ and $I_i^{+}I_j^{+}$ are dropped. 
When both of the two approximations are applied, the only decoherence mechanism left is the secular spin
flip-flop described in Section~\ref{SubSubSect:Secular}. 
Through the comparison, the importance of various types of interaction is identified.

\subsection{Free-induction decay}
\label{subsect:FID}

In FID experiments, the coupling to the host $^{14}$N nuclear spin induces fast oscillations ($\sim \text{MHz}$) on the
decoherence envelope induced by the $^{13}$C bath spins. 
To single out the effect due to the $^{13}$C bath spins, only the envelope is presented here.

\begin{figure}[tb]
  \includegraphics[width=\columnwidth]{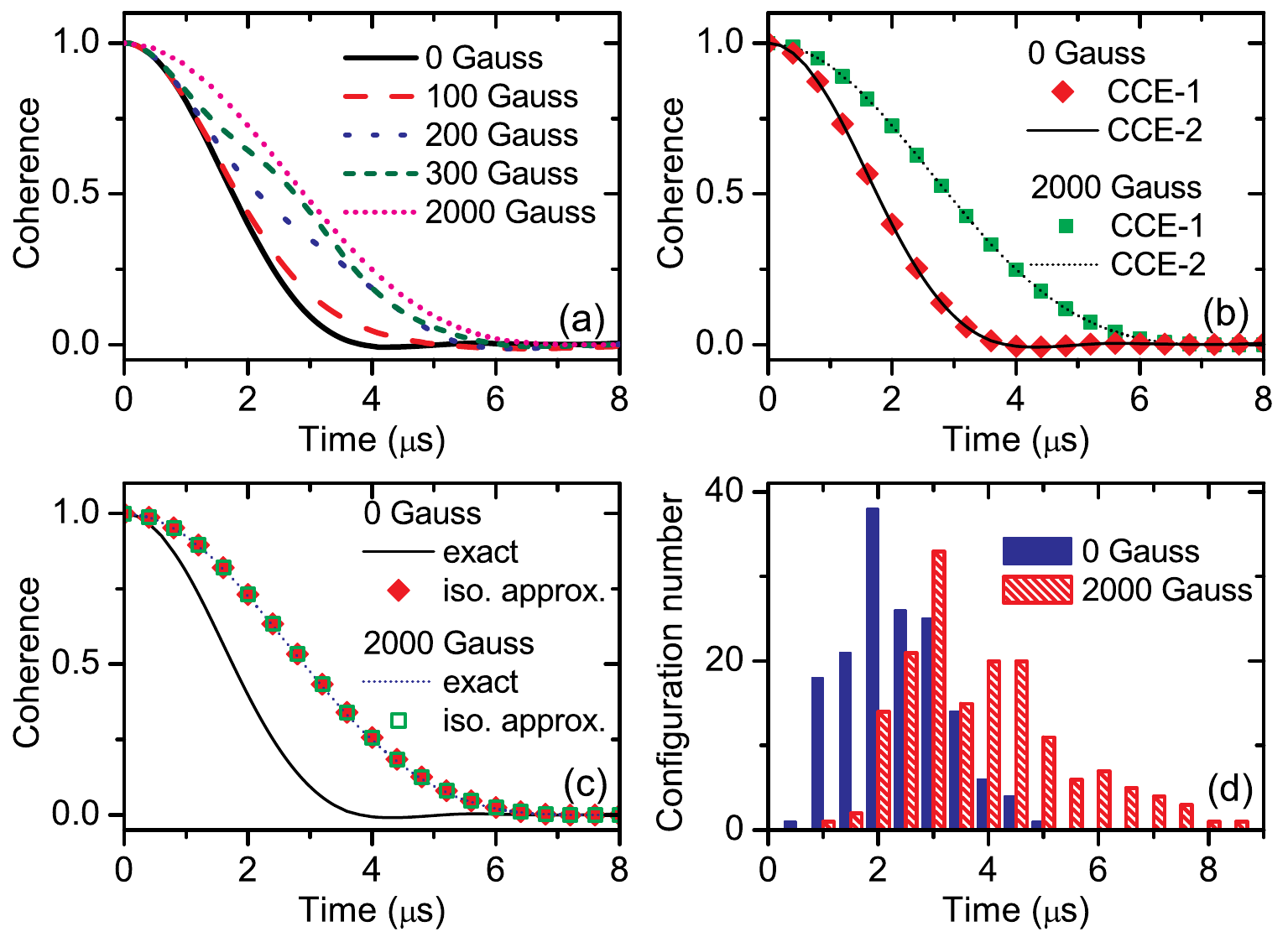}
  \caption{(a) FID of an NV center in a typical nuclear spin configuration under various magnetic field strength.
  (b) The contributions to FID of different cluster sizes (CCE orders).
  (c) The exact FID coherence and the coherence obtained by applying the isotropic approximation in zero and
  strong fields.
  (d) Histogram of FID coherence time $T_2^{*}$ distribution under zero and strong fields.
  }
  \label{FIG:FID}
\end{figure}

Figure~\ref{FIG:FID}(a) shows the FID of an NV center spin in a typical $^{13}$C bath spin configuration under
various magnetic field strength. 
The coherence decays in several $\mu \text{s}$. During such a short time, the dipolar interaction between nuclear spins
(typically $ \lesssim \text{kHz}$) is too weak to take effect. Thus, the FID is caused by the non-interacting nuclear
spins, and the coherence is the product of the contributions of each single spins, i.e. $L_{\text{FID}}(t)= \prod_j
L_{j,\text{FID}}(t)$. This is confirmed in Fig.~\ref{FIG:FID}(b), where the CCE-1 results are convergent.
Taking the pair correlation into account (i.e. CCE-2) does not change the coherence. 

The FID decoherence behavior depends on the magnetic field strength.
The coherence time in strong field is longer than that in weak or zero field, and the decoherence profile deviates
from the Gaussian decay in the medium field [e.g., $B=200$~Gauss, see Fig.~\ref{FIG:FID}(a)].

It is the anisotropic nature of the hyperfine coupling in NV centers that causes the distinguished decoherence features.
In Fig.~\ref{FIG:FID}(c), we compare the exact FID coherence results with the results obtained with the isotropic
approximation. 
In zero magnetic field, the anisotropic terms of the hyperfine coupling play an important role. If
the anisotropic terms are ignored in calculations, the FID decoherence behavior in zero field will be identical to that
in the strong field limit. While in the strong field case (e.g., $B=2000$~Gauss), the contribution of the anisotropic
terms is negligible.

This decoherence behavior can be understood by analyzing the single nuclear spin dynamics.
For each single spin, the contribution to the FID is shown in Eq.~\ref{Eq:SingleSpinCoherenceHahnEcho}(a). In the weak
field ($B\lesssim 1$~Gauss), the precession angle about the external field $\theta_j^{(0)}\approx 0$. According to Eq.~\ref{Eq:SingleSpinCoherenceHahnEcho}(a), the coherence is
\begin{equation}
L_{\text{FID}}(t)\approx \prod_j \cos\frac{A_j^{(+1)}t}{2}\approx e^{-\Gamma^2t^2/2},
\end{equation}
where $A_j^{(+1)}$ is the magnitude of the hyperfine field of the $j$th nuclear spin with electron spin in $|+1\rangle$
state and $\Gamma^2=\sum_j 
\left(A_j^{(+1)}\right)^2/4$ is the FID decay rate in zero or weak fields. 
In the strong field limit ($B\gg 300$~Gauss), the effective fields $\mathbf{h}_j^{(0)}$ and $\mathbf{h}_j^{(+1)}$
are almost parallel with each other, i.e. $\varphi_j\approx 0$ [see Fig.~\ref{FIG:singlespin}]. 
In this case, the coherence is
\begin{align}
L_{\text{FID}}(t)\approx \prod_j \cos\frac{\theta_j^{(+1)}-\theta_j^{(0)}}{2}\approx \prod_j
\cos\frac{A_{j,z}^{(+1)}t}{2} \approx e^{-\Gamma_z^2t^2/2},
\end{align}
where $A_{j,z}^{(+1)}$ is the hyperfine field component along the magnetic field direction, and $\Gamma_z^2=\sum_j 
\left(A_{j,z}^{(+1)}\right)^2/4$ is the FID decoherence rate in strong filed limit. 

Since $A_{j,z}^{(+1)}\le A_{j}^{(+1)}$ always holds, the FID decoherence time in the strong field is longer than that in
the weak field. 
Physically, this is because, in the strong field limit, the $^{13}$C nuclear spins are frozen along the quantization
axis defined by the applied magnetic field. Thermal fluctuations enable the nuclear spins pointing parallel or anti-parallel
the magnetic field direction, providing an Overhauser filed proportional to the hyperfine fields components
$A_{j,z}^{(+1)}$. While in the zero field limit, the quantization axis of each nuclear spins can be chosen along the
hyperfine field directions. 
The nuclear spins are parallel or anti-parallel along the hyperfine field, so that the Overhauser field strength is
proportional to the magnitude $A_j^{(+1)}$ of the hyperfine fields. 
The stronger Overhauser fields in weak magnetic field induces the faster FID decoherence.
In Fig.~\ref{FIG:FID}(d), we show the histogram of FID coherence time for about 200 randomly generated bath spin
configurations (the random positions of $1.1\%$ $^{13}$C in the diamond lattice) under strong and zero magnetic
fields. The FID decoherence time spreads over several microseconds for different spin configurations. 
The mean decoherence time over different configurations under zero and strong magnetic field is
$T_2^{*}\equiv\sqrt 2 / \Gamma =2.1$~$\rm {\mu s}$ and $T_{2,z}^{*}\equiv\sqrt 2 / \Gamma_z =3.6$~$\rm {\mu s}$,
respectively.

\begin{figure*}[tb]
  \includegraphics[width=0.95\textwidth]{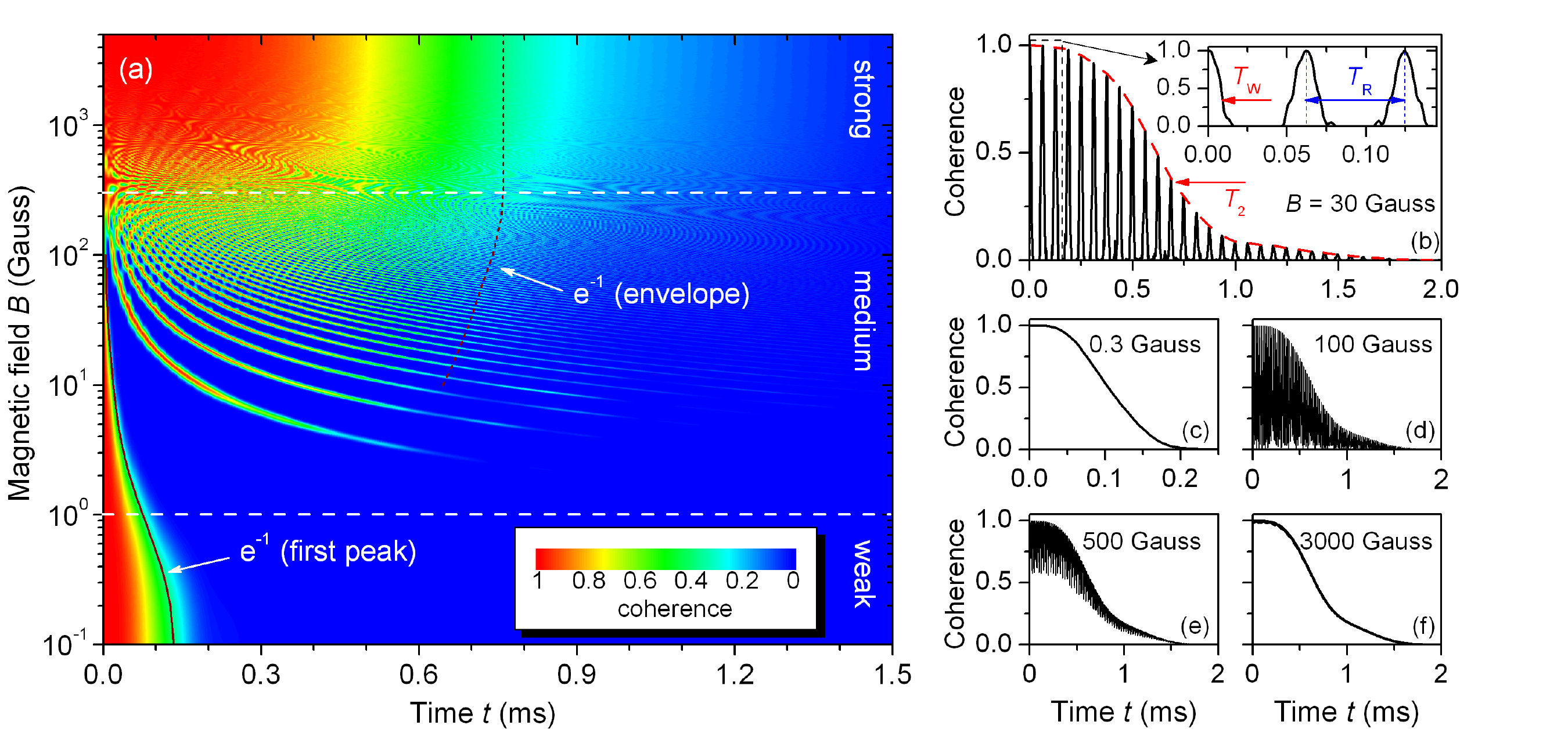}
  \caption{(a) Hahn echo of NV center spin coherence between $|0\rangle$ and $|+1\rangle$ as a function of total evolution time
  $t$ and magnetic field $B$. The strong, medium and weak magnetic field regimes, in which the NV center spin has different
  decoherence behavior, are separated by the horizontal lines. The solid (dashed) line indicates the time at which the
  NV center spin coherence decays to the $1/e$ value for the first peak (envelope), i.e. the time $T_{\text{W}}$ ($T_2$) defined in (b). 
  (b)-(f) Cross-sections of (a) for various magnetic field strength. 
  (b) The collapse and revival effect under $30$~Gauss magnetic field with three different timescales indicated. 
  The red dashed line is the envelope of the revival peaks, which decays to $1/e$ at time $T_2$.
  Inset: close-up of the first several peaks. The half peak width $T_{\text{W}}$ and the revival period $T_{\text{R}}$
  are indicated.
  (c) In a weak magnetic field of 0.3~Gauss (approximately the earth magnetic field strength), the NV center spin
  coherence decays in a timescale $T_{\text{W}}$, and does not revive. 
  (d)-(f) The suppression of the collapse and revival effect by increasing the magnetic field strength.
   For $B=100$~Gauss in (d), the revival peaks get closer to each other. The coherence does not collapse to zero and
   the revival peaks are not well resolved under $B=500$~Gauss in (e). 
   For $B=3000$~Gauss in (f), the collapse and revival effect is completely suppressed. }
  \label{FIG:HahnEcho3D}
\end{figure*}

In a magnetic field between the strong and weak field limit ($1\text{~Gauss}<B\lesssim 300 \text{~Gauss}$), as shown in
Fig.~\ref{FIG:FID}(a), the FID decoherence profile deviates from the Gaussian decay. 
This decoherence behavior is caused by the quantum fluctuation induced by the $^{13}$C spins close to the NV center. 
In a medium field, the Zeeman energy is comparable to the hyperfine
coupling strength for the closely located $^{13}$C spins. In this case, as shown in Fig.~\ref{FIG:singlespin}, the
magnitudes of effective fields $\mathbf{h}_j^{(0)}$ and $\mathbf{h}_j^{(+1)}$ are comparable and they are, in general,
not parallel. Thus, the eigenstates of the noise operator $b_z$ (i.e. the nuclear spin state has definite
Overhauser field) are not eigenstates of the total Hamiltonian. 
In this way, the quantum fluctuation arises from the superposition of the nuclear
spin states with different Overhauser fields, and causes the non-Gaussian decay in the medium magnetic field. 
This distinguished feature of FID decoherence is essentially attributed to the anisotropic nature of the hyperfine
coupling of NV centers. In QDs and shallow donors, where the isotropic hyperfine coupling dominates, the
external field and the Overhauser field are always parallel, and such non-Gaussian FID would be absent.

\subsection{Hahn echo}
\label{subsect:hahn}

\begin{figure}[tb]
  \includegraphics[width= \columnwidth]{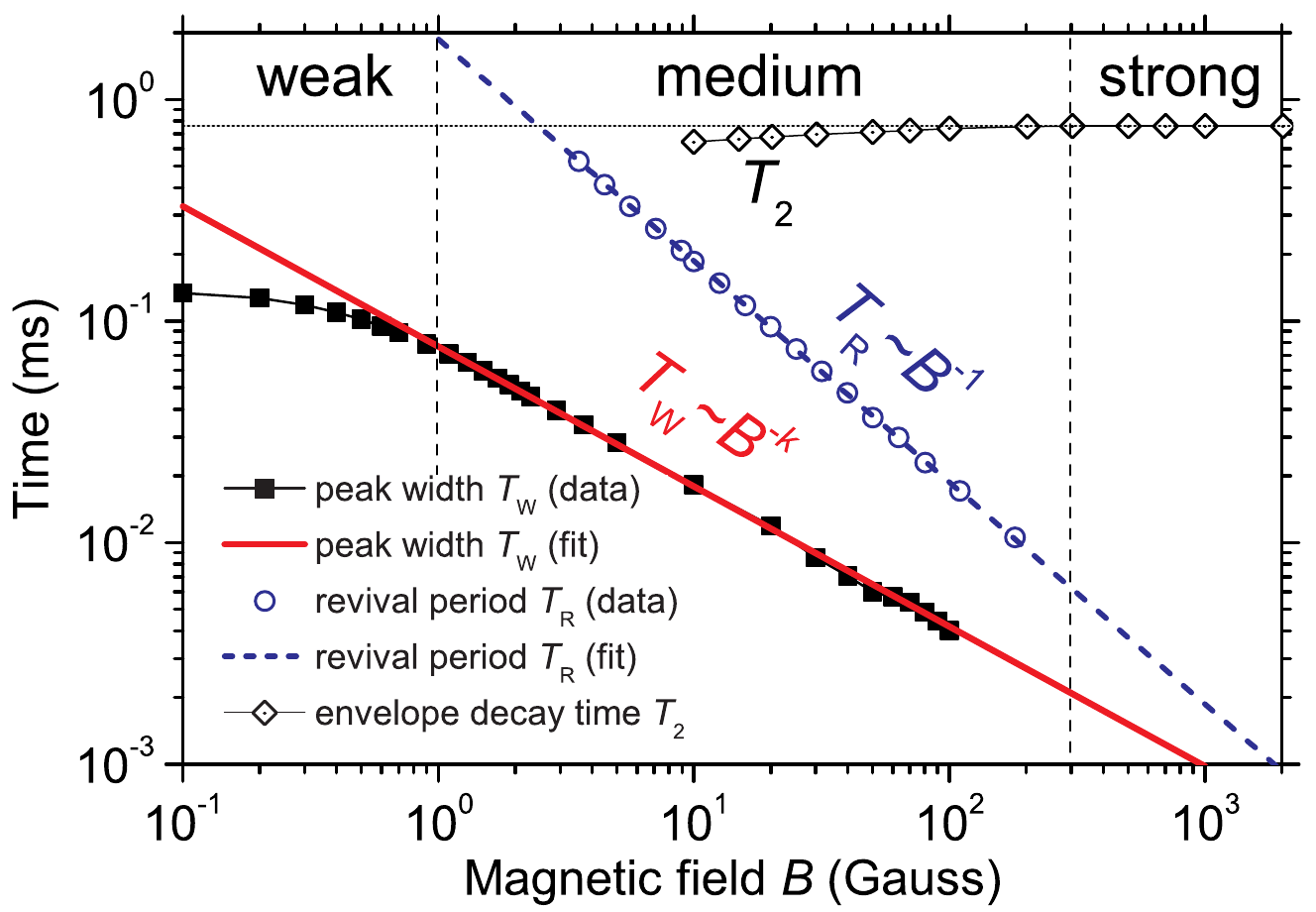}
  \caption{The magnetic field dependence of $T_{\text{W}}$, $T_{\text{R}}$ and $T_{2}$. The black
  line with square symbols is the calculated peak width, and the red solid line is obtained by fitting the data in the
  region $1~\text{Gauss}<B<100~\text{Gauss}$. The peak width $T_{\text{W}}$ depends on $B$ as $B^{-k}$ with
  $k=0.63$. The value of $k$ slightly depends on the specific nuclear spin positions around the NV center.
  For $B< 1$~Gauss, the peak width $T_{\text{w}}$ deviates from the $B^{-k}$ law, which defines the weak magnetic field
  regime.
  The revival time $T_{\text{R}}$ is inversely proportional to the magnetic field strength $B$ (circle symbols and blue
  dashed line). The rough boundary between the medium and strong magnetic field regimes is defined by the approaching of
  $T_{\text{R}}$ to $T_{\text{W}}$ about $B= 300$~Gauss. 
  The envelope decay time $T_2$ (black line with diamond symbols) in the strong magnetic field ($>300$~Gauss) is
  constant, and decreases as the magnetic field is decreased to $\sim 10$~Gauss. For $B<10$~Gauss, there are too few revival peaks
  to determine the envelope decay time, and $T_W$ should be taken as the relevant decoherence time. }
  \label{FIG:CoherenceTime}
\end{figure}

The thermal noise is eliminated by applying the Hahn echo. 
The decay of the spin echo is caused by the dynamical fluctuations of the nuclear spin bath.
The decoherence time of a certain NV center spin depends on the bath spin configurations around it.\cite{Maze2008a} 
In Fig.~\ref{FIG:HahnEcho3D}(a), we show a typical Hahn echo coherence as a function of the total
evolution time and the magnetic field strength.

Three magnetic field regimes, in which the NV center spin coherence decays in different manners, are roughly
distinguished by the horizontal lines in Fig.~\ref{FIG:HahnEcho3D}(a). 
In the weak field regime ($B\lesssim 1$~Gauss), the coherence decays monotonically within $\sim 200$~$\mu \text{s}$ [see
Fig.~\ref{FIG:HahnEcho3D}(c)].
In the medium field regime ($1\text{~Gauss}\lesssim B \ll 300 \text{~Gauss}$), the main feature is the periodic
coherence collapse and revival under an overall decay envelope [see Fig.~\ref{FIG:HahnEcho3D}(b) and (d)].
In the strong field regime ($B > 300 \text{~Gauss}$), the collapse and revival effect is greatly suppressed
[Fig.~\ref{FIG:HahnEcho3D}(e)] and finally vanished [Fig.~\ref{FIG:HahnEcho3D}(f)]. 
The coherence monotonically decays as in the weak field regime, but with a longer timescale of $\sim 700$~$\mu
\text{s}$.

To quantitatively characterize the decoherence behavior, three timescales are defined [see Fig.~\ref{FIG:HahnEcho3D}(b)]: 
(i) the electron spin coherence collapses and revives within a period $T_{\text{R}}$, forming a periodic peak
structure,
(ii) for each coherence peak, the half width is denoted as $T_{\text{W}}$ (the half width at $1/e$ height,
approximately the same for all the peaks),
and (iii) the overall envelope decays to $1/e$ in a slower timescale $T_2\sim \rm ms$.
The peak decay time $T_{\text{W}}$ of the first (half) peak and the envelope decay time $T_2$ are highlighted in
Fig.~\ref{FIG:HahnEcho3D}(a) with the solid and dashed contours, respectively.

The field dependence of $T_{\text{R}}$, $T_{\text{W}}$ and $T_2$ are summarized in Fig.~\ref{FIG:CoherenceTime}.
The revival period $T_{\text{R}}$ is inversely proportional to the magnetic field strength as expected. The envelope
decay time $T_2$ increases with increasing the magnetic field in the medium field regime, and becomes constant in the
strong field regime. The peak decay time $T_{\text{W}}$ decreases with decreasing the magnetic field as $T_{\text{W}}\propto B^{-k}$
with $k\approx 0.6$ in the medium field regime, and approaches the revival period $T_{\text{R}}$ when the field is
close to the boundary of the strong field regime. In the weak field regime, $T_{\text{W}}$ deviates from the $B^{-k}$
law, and is saturated in the zero field limit. The magnetic field regimes are divided according to not only the
apparent field dependence of $T_{\text{R}}$, $T_{\text{W}}$ and $T_2$ shown in Fig.~\ref{FIG:CoherenceTime}, 
but also the different underlying decoherence mechanisms, which will be discussed below.

\subsubsection{Medium magnetic field regime}

\begin{figure}[tb]
  \includegraphics[width= \columnwidth]{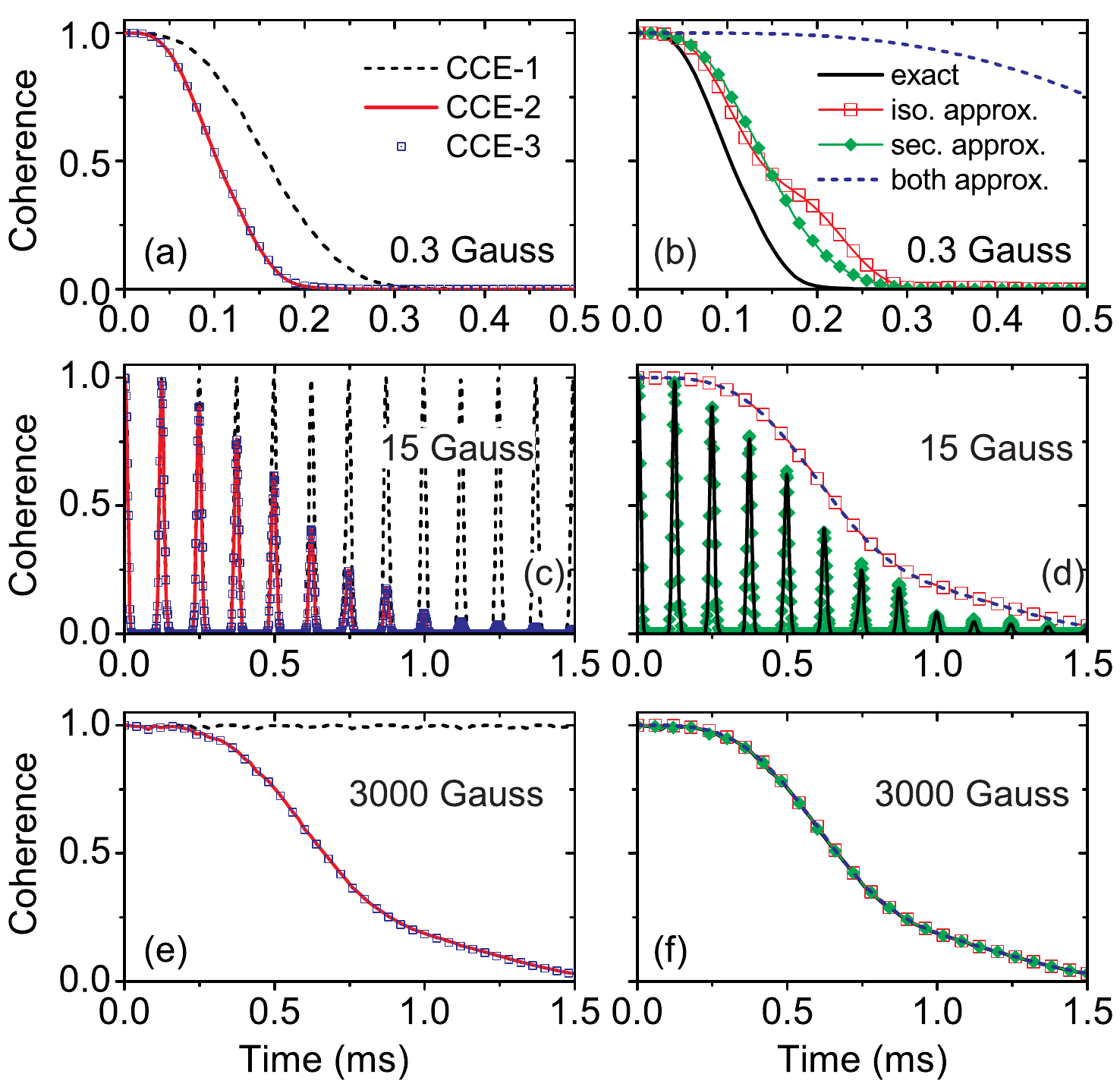}
  \caption{Decoherence mechanisms of Hahn echo in different magnetic field regimes. 
  (a) The contributions to decoherence of different cluster sizes (CCE orders) in $B=0.3$~Gauss.
  (b) The contributions to decoherence of different types of interaction in $B=0.3$~Gauss. 
  The red line with square symbols is the coherence calculated with the isotropic approximation
  of the hyperfine interaction. 
  The green line with diamond symbols is the coherence calculated with the secular approximation. 
  The blue dashed line is the coherence obtained with both isotropic and secular approximations. 
  The black solid line indicates the exact results without approximations. 
  (c) and (d) are the same as (a) and (b), respectively, but for $B=15$~Gauss.  
  (e) and (f) are the same as (a) and (b), respectively, but for $B=3000$~Gauss. }
  \label{FIG:Hahn2}
\end{figure}

Let us start from the medium field regime with the magnetic field $1\text{~Gauss}\lesssim B < 300 \text{~Gauss}$. 
The contribution to decoherence of different cluster sizes is shown in Fig.~\ref{FIG:Hahn2}(c). 
As discussed in Section~\ref{Sect:spin_process_pic}, the collapse of the electron spin coherence is induced by the single
$^{13}$C nuclear spin precession. 
This is confirmed in Fig.~\ref{FIG:Hahn2}(c), where the CCE-1 gives rise to the periodically collapse and revival.
The coherence would recover perfectly if only the single-spin clusters are considered. 
The envelope decays when the two-spin clusters are taken into account (CCE-2).
This means the envelope decay is caused by nuclear spin pair dynamics. 
The higher order correlations of more than two nuclear spins are not important in Hahn echo since the CCE-2 gives
the convergent result.

Furthermore, we identify the contributions to decoherence of different types of interaction.
As shown in Fig.~\ref{FIG:Hahn2}(d), the secular approximation gives almost the same result as the exact one. 
This implies the non-secular spin flipping has negligible effect on spin decoherence in the medium field regime. 
It is understandable since the dipolar interaction strength ($\ll $~kHz) is much smaller than the Zeeman energy in this
field regime ($>$~kHz).
Since the anisotropic coupling of the hyperfine interaction is responsible to the collapse and revival effect as
discussed in Section~\ref{SubSect:SingleSpin}, the coherence collapse and revival do not occur when the isotropic
approximation is applied. 
In addition, the anisotropic terms results in a shorter envelope decay time $T_2$ than the coherence time with only the
secular flip-flop process taking into account [i.e. with both secular and isotropic approximation applied, see the blue
dashed line Fig.~\ref{FIG:Hahn2}(d)].

The effect of the anisotropic terms on the coherence envelope is explained as follows. 
Notice that, besides the secular flip-flop process
$|\uparrow\downarrow\rangle\leftrightarrow|\downarrow\uparrow\rangle$, the energy non-conserving two-spin flipping
processes, such as $|\uparrow\uparrow\rangle \leftrightarrow|\downarrow\uparrow\rangle$, are induced by the
transverse hyperfine field $A_{j,x}$ or $A_{j,y}$.
In the medium field regime, these energy non-conserving processes can occur since the hyperfine interaction strength
for the nuclear pair close to the NV center ($>10$~kHz) could be stronger than or comparable to the nuclear spin Zeeman
energy. 
With such kind of processes activated, the envelope decay in the medium field regime is faster than the decoherence
solely caused by the secular nuclear spin flip-flop process [the blue line in Fig.~\ref{FIG:Hahn2}(d)]. 
With increasing the magnetic field strength, this kind of processes are gradually suppressed due to the increasing
Zeeman energy. As a result, the envelope decay time $T_2$ [the dashed contour in Fig.~\ref{FIG:HahnEcho3D}(a) and the diamond symbols
in Fig.~\ref{FIG:CoherenceTime}] increases and approaches the value in the strong field regime.

As for the single peak decay time $T_{\text{W}}$, we plot the magnetic field dependence of $T_{\text{W}}$ in
Fig.~\ref{FIG:CoherenceTime} in comparison with the revival period $T_{\text{R}}$. 
According to Eq.~(\ref{Eq:SingleSpinCoherenceHahnEcho}), in the short time limit, the electron spin coherence is
expressed as
\begin{equation}
L(t)\approx \prod_j\left(1-\frac{1}{8}\left[(A_{j,x}^{(+1)})^2+(A_{j,y}^{(+1)})^2\right] B^2\tau^4\right),
\end{equation}
which implies the decay time of a single peak scales with the magnetic field strength as $\sim B^{-1/2}$.
The numerical result gives the peak width $T_{\text{W}}\propto B^{-k}$ with the index $k\approx 0.6$.
The deviation of index $k$ from $1/2$ is due to the violation of the short-time condition for the $^{13}$C spins close
to the NV center.

\subsubsection{Weak magnetic field regime}

The revival period $T_{\text{R}}$ increases as the  magnetic field is decreased. 
As shown in Fig.~\ref{FIG:HahnEcho3D}, for $B\lesssim 2$~Gauss, the subsequent revival peaks are hardly observed due
to the decay of the envelope, and only the first (half) peak is relevant in this case. 
Meanwhile, for $B < 1$~Gauss, the dependence of $T_{\text{W}}$ on $B$ begins to deviate from the $~B^{-k}$ law as
governed in the medium magnetic field regime, and becomes saturated approaching the zero field
(see Fig.~\ref{FIG:CoherenceTime}). Thus, $B = 1$~Gauss defines the rough boundary of the weak field regime.  
In this regime, the decoherence time is defined by $T_W$.

To analyze the decoherence mechanism in the weak field regime, in Fig.~\ref{FIG:Hahn2}(a), we check the
CCE convergence to find out the relevant clusters responsible to the decoherence.
Being different from the medium field regime, where the single spin dynamics dominates the peak decay, 
the single spin precession process becomes less important in the weak field regime (see
Section~\ref{SubSect:SingleSpin}).
As shown in Fig.~\ref{FIG:Hahn2}(a), under a magnetic field $B=0.3$~Gauss, the single spin precession (CCE-1)
contributes a part of the decoherence, and the coherence converges when the pair dynamics (CCE-2) is taken into account.

In Fig.~\ref{FIG:Hahn2}(b), different types of interaction in the weak field regime are analyzed.
In this regime, the conservation of Zeeman energy is not a necessary requirement. 
Thus, the Zeeman energy non-conserving spin flipping processes induced by anisotropic hyperfine coupling and the
non-secular terms in the dipolar interaction are activated. 
As discussed in Section~\ref{Sect:spin_process_pic}, the noise amplitudes produced by these processes are typically
stronger than that produced by the secular flip-flop process of nuclear spin pair.
In this way, the  energy non-conserving spin flipping processes become the dominant decoherence mechanism [see the red and blue lines with symbols in Fig.~\ref{FIG:Hahn2}(b)], while the secular flip-flop process has
negligible contribution to the decoherence [see the blue dashed line in Fig.~\ref{FIG:Hahn2}(b)].

\subsubsection{Strong magnetic field regime}

Since $T_{\text{W}}$ decreases slower than $T_{\text{R}}$, as shown in Fig.~\ref{FIG:CoherenceTime},
in strong magnetic fields ($B \gtrsim 300$~Gauss), $T_{\text{R}}$ and $T_{\text{W}}$ are of the same order. 
The collapse and revival effect due to the single nuclear spin precession is greatly suppressed in the strong field
regime, and the peak structure will finally disappear in the strong field limit [see Fig.~\ref{FIG:HahnEcho3D}(d)-(f)]. 

In Fig.~\ref{FIG:Hahn2}(e), the CCE convergence is shown. In strong magnetic field, the nuclear spins are frozen along
the magnetic field direction, and the single spin precession (CCE-1) is almost completely suppressed. 
The decoherence is caused by the nuclear spin pair flip-flops (CCE-2).
Furthermore, all the energy non-conserving two-spin processes are suppressed by the strong
field, leaving the secular flip-flop process being the only decoherence mechanism in the strong field regime. 
This is confirmed in Fig.~\ref{FIG:Hahn2}(f), where neglecting the anisotropic coupling and the non-secular terms does
not affect the decoherence behavior. 
Since the absolute magnitude of Zeeman energy is not involved in the secular flip-flop process, the coherence time $T_2$
does not depends on the magnetic field strength anymore in the strong field regime [see the dashed contour in
Fig.~\ref{FIG:HahnEcho3D} and the diamond symbols in Fig.~\ref{FIG:CoherenceTime}].

\subsection{Multi-pulse dynamical decoupling controls}
\label{SubSec:Multi-pulse}
In this Section, we study the effects of different decoherence mechanisms in different magnetic field regimes, for NV
center spins under multi-pulse DD control.

\subsubsection{Medium magnetic field regime}

\begin{figure}[tb]
  \includegraphics[width= \columnwidth]{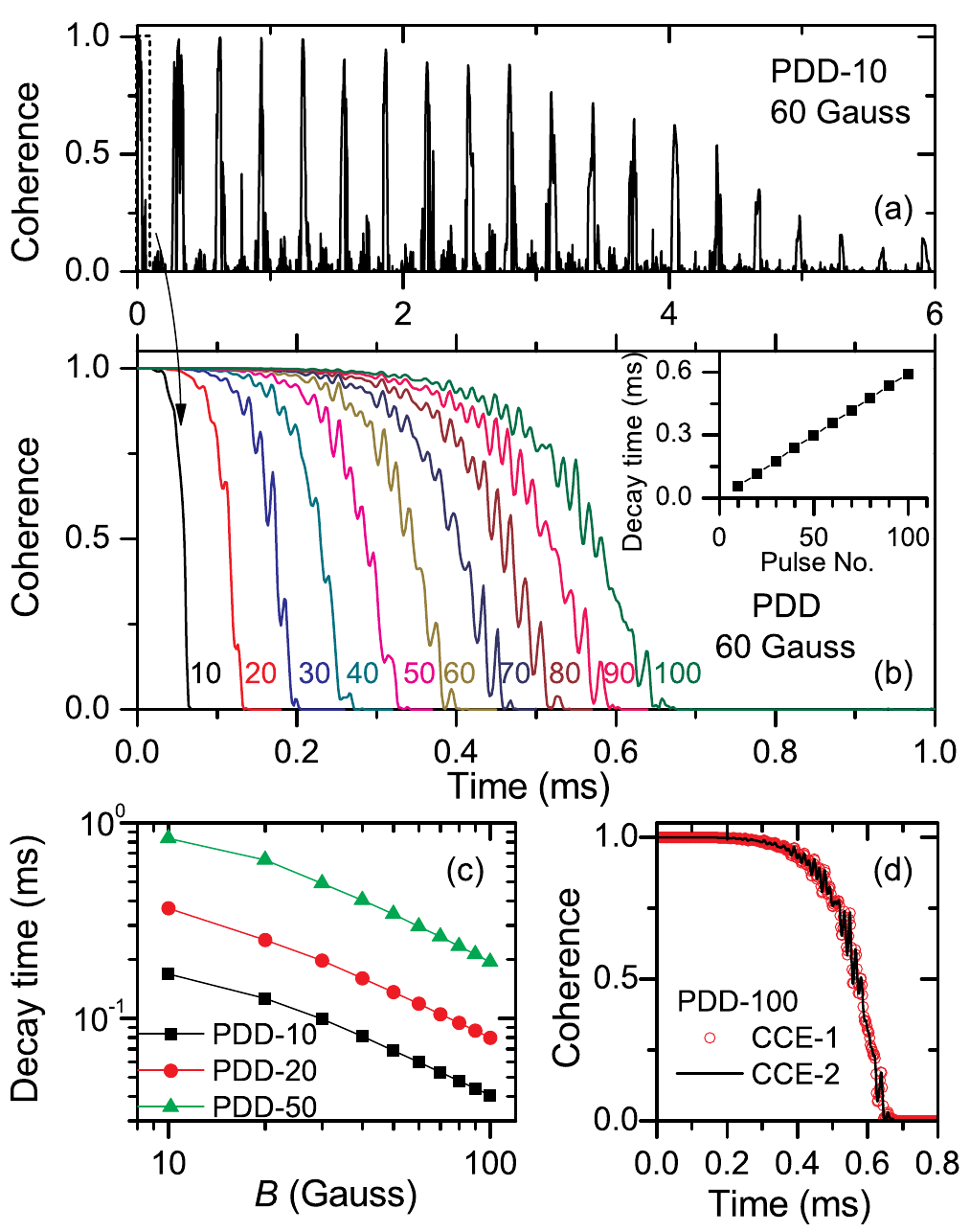}
  \caption{(a) Electron spin coherence between $|0\rangle$ and $|+1\rangle$ under PDD-10 control in
  the medium magnetic field regime ($B=60$~Gauss). Similar collapse and revival effect is observed, but with a
  longer timescale than the Hahn echo case. 
  (b) The decoherence within the first (half) peak under PDD controls of various pulse numbers indicated by the
  integers associated with each curves. For the sake of clarity, the subsequent revivals are not shown. 
  The inset shows that the decay time, at which the coherence decays to $1/e$ value, increases linearly with increasing
  pulse number.
  (c) The magnetic field dependence of the decay time (at $1/e$ height) for various PDD pulse numbers.
  (d) The CCE convergence of the PDD-100 result under $B=60$~Gauss.
  The single spin dynamics (CCE-1) is the dominating decoherence mechanism in this magnetic field regime.
  \label{FIG:fig8_MediumField}
  }
\end{figure}

Let us first consider the medium magnetic field regime, in which the spin coherence under DD controls has been studied
experimentally.\cite{Ryan2010} 
As a direct generalization of the Hahn echo, the electron spin coherence under PDD control also presents the
collapse and revival effect [see Fig.~\ref{FIG:fig8_MediumField}(a)].
We will first focus on the dynamics within a revival peak, and investigate the DD control effect on the peak decay time.

The electron spin coherence under magnetic field $B=60$~Gauss and multi-pulse PDD controls is
shown in Fig.~\ref{FIG:fig8_MediumField}(b). The peak decay time $T_{\text{W}}$ is extended from $\sim 10$~$\rm{\mu s}$
for the Hahn echo to $\sim 600$~$\rm{\mu s}$ for the 100-pulse PDD control. 
The small oscillations on the decoherence curves arise from some
$^{13}$C nuclear spins located relatively close to the NV center which have strong hyperfine interaction ($A_j^{(+1)}\sim 100$~kHz). 
The coherence time increases linearly with increasing the DD control pulse number [see the inset of
Fig.~\ref{FIG:fig8_MediumField}(b)], which is consistent with the experimental observations in the same magnetic field
regime.\cite{Ryan2010} 
In contrast, the experiment of NV center coherence in electron spin baths shows that the coherence time depends on the
DD control pulse number $N$ as $ N^{2/3}$.\cite{Lange2010} 
This is because the different features of the noise spectra in nuclear and electron spin baths.
In electron spin baths, the noise spectrum is of Lorentzian shape with a long high-frequency tail, while the noise
spectrum due to the nuclear spin bath has a hard high-frequency cut-off. 
In particular, in the medium magnetic field regime, the noise spectrum is a peak located around the Lammor frequency.
The high frequency noises in electron spin baths make the DD less efficient in protecting the NV center spin coherence
(i.e. the coherence time increases slower with increasing pulse number). 
The quantitative relation between the coherence and noise spectrum is an interesting problem but beyond the scope of
this paper. It will be discussed elsewhere.

\begin{figure}[tb]
  \includegraphics[width= \columnwidth]{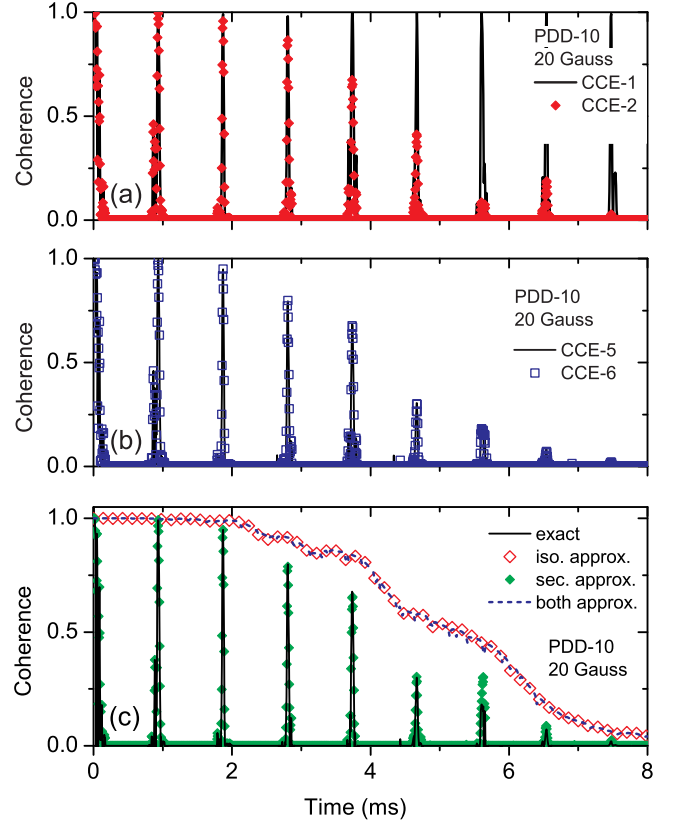}
  \caption{(a) and (b) The contributions to overall decoherence of different cluster sizes (CCE orders) in $B=20$~Gauss and
  PDD-10 control. CCE-1 give rise to the collapse and revival effect. The envelope decay is caused mainly by CCE-2. The
  calculation is converged at CCE-6. 
  (c) The contributions to decoherence of different types of interaction in $B=20$~Gauss and PDD-10 control.
  \label{FIG:fig8_MediumField1}
  }
\end{figure}

As shown in Fig.~\ref{FIG:fig8_MediumField}(c), the peak decay time under multi-pulse PDD controls has the similar
magnetic field dependence as in the Hahn echo case. 
The similar magnetic field dependence of the peak width suggests that the same decoherence mechanism, i.e. the single
spin precession process, is responsible for the single peak decay under multi-pulse PDD controls.
To confirm this point, the CCE convergence for the 100-pulse PDD control is shown in
Fig.~\ref{FIG:fig8_MediumField}(d). 
It is observed that CCE-1 almost gives the convergent coherence. Including the correlations between more spins induces
negligible change to the coherence result. 
This convergence indicates that, although the coherence time under
multi-pulse DD is prolonged by more than one order of magnitude as compared with the Hahn echo, the dominating mechanism for the initial coherence decay is still
the single spin precession. 

In Fig.~\ref{FIG:fig8_MediumField1}(a) and (b), we check the CCE convergence in a longer timescale to study the envelope
decay. Similar to the Hahn echo case, the single-spin precession (CCE-1) results in the perfect revival, while the
nuclear spin pair correlation (CCE-2) contributes the main part of the envelope decay. 
As the coherence is prolonged under DD control, the higher order correlations (clusters with more than two spins) begin
to take effect on the envelope decay and give a small correction to the CCE-2 result. 
As shown in Fig.~\ref{FIG:fig8_MediumField1}(b), the calculation converges with the correlations of six-spin clusters
included (CCE-6).

The contribution of different type of interaction under DD control is shown in Fig.~\ref{FIG:fig8_MediumField1}(c). 
The conclusion is similar to the Hahn echo case. 
In the medium field regime, the non-secular spin flipping contributes little to the decoherence. 
The anisotropic terms of hyperfine coupling cause the collapse and revival phenomenon and provide the additional
envelope decay besides the one caused by the secular spin flip-flop process. 

\subsubsection{Weak magnetic field regime}
\label{SubSubSect:WeakField}

In the weak magnetic field regime ($B\lesssim1$~Gauss), the two electron spin states $|\pm 1\rangle$  are hardly
resolved through the transition frequencies. 
In this case, instead of the coherence between $|0\rangle$ and $|\pm 1\rangle$ states, we investigate the coherence
between $|+1\rangle$ and $|-1\rangle$, which can be generated and manipulated in experiment by linearly polarized
microwave pulses.\cite{Note1,Zhao2011a}

\begin{figure}[tb]
  \includegraphics[width= \columnwidth]{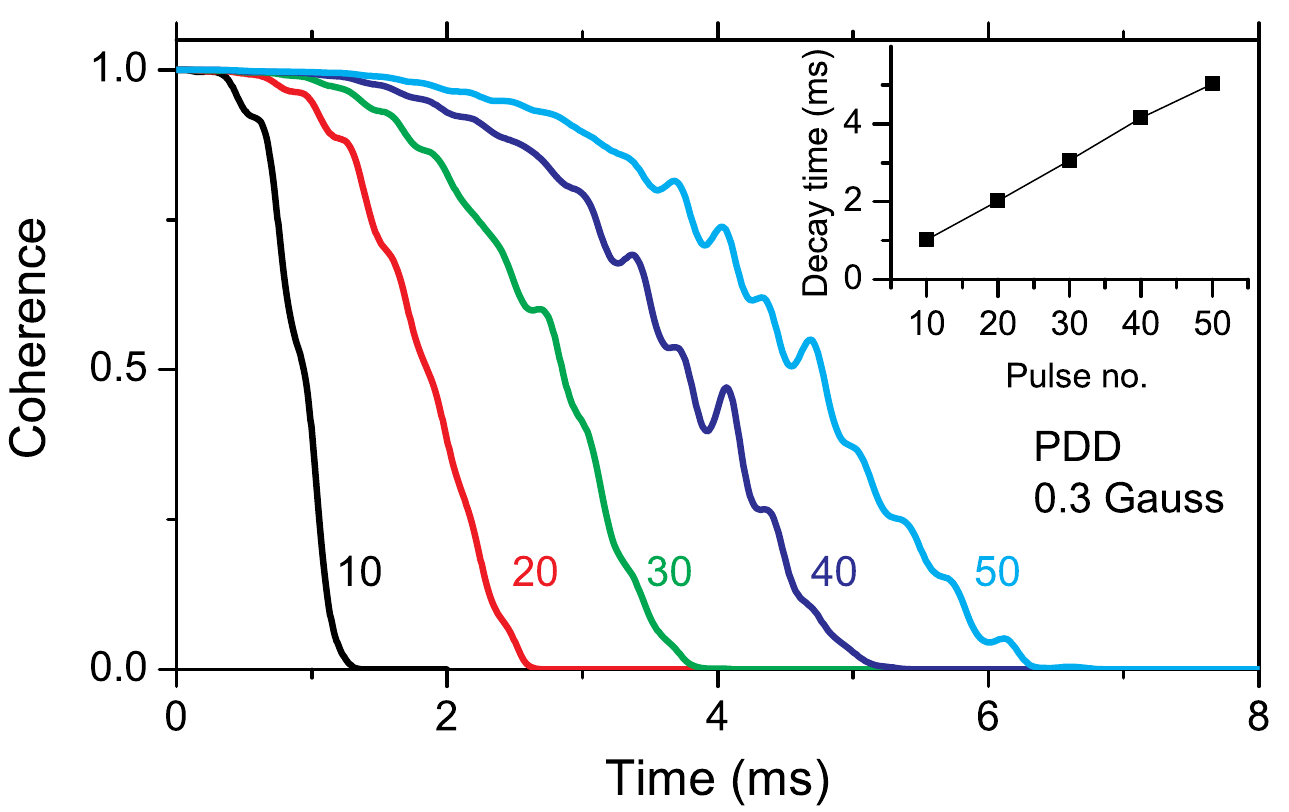}
  \caption{Electron spin coherence between $|+1\rangle$ and $|-1\rangle$ under weak magnetic field (0.3~Gauss) and PDD controls. 
  The integer numbers associated with each curves are the pulse numbers of the PDD controls. 
  Inset: the coherence time $T_{\text{W}}$ (at $1/e$ height), increases linearly with increasing pulse number.
  }
  \label{FIG:fig9_DDzeroB}
\end{figure}

The electron spin coherence under $B=0.3$~Gauss and multi-pulse PDD
controls is presented in Fig.~\ref{FIG:fig9_DDzeroB}. 
Similar to the medium field case, the coherence time $T_{\text{W}}$ increases linearly with increasing pulse number
(see the inset of Fig.~\ref{FIG:fig9_DDzeroB}).
 
\begin{figure}[tb]
  \includegraphics[width= \columnwidth]{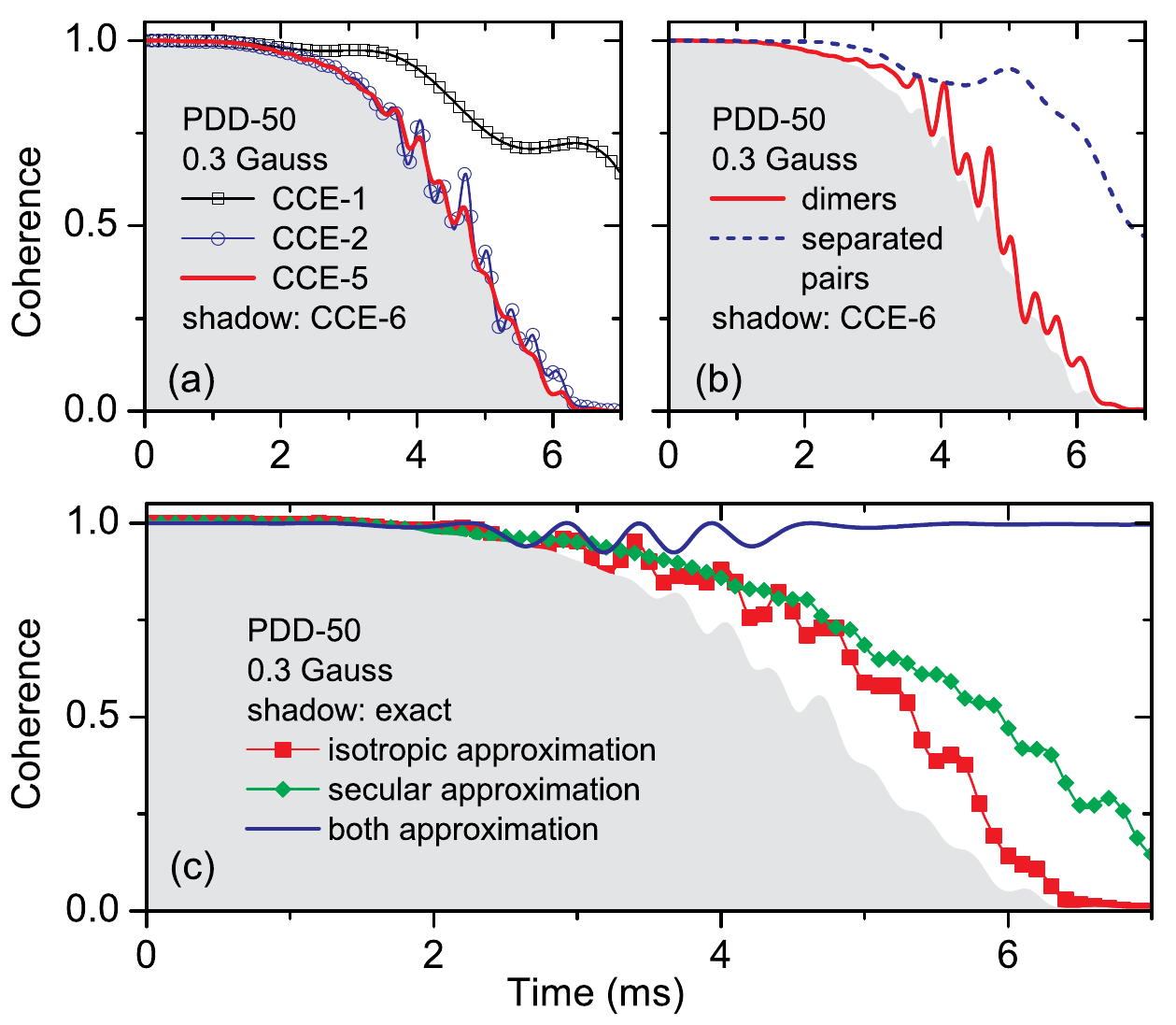}
  \caption{(a) The CCE convergence for PDD-50 under weak magnetic field $B=0.3$~Gauss. The 2-spin clusters give the dominating
  contribution (CCE-2), and the result converges by including up to 6-spin clusters (CCE-6). 
  (b) Contributions to NV center spin decoherence from $\sim 12$ nuclear spin dimers (red solid line) and the
  large number of separated pairs (blue dashed line) in the bath. The decoherence in the weak field regime is mainly
  caused by the dimers. (c) Decoherence caused by different types of interaction in $B=0.3$~Gauss and PDD-50 control.
  }
  \label{FIG:DDzeroB2}
\end{figure}

To analyze the decoherence mechanisms, we examine the CCE convergence. Taking the
coherence under PDD-50 control for example [Fig.~\ref{FIG:DDzeroB2}(a)], the CCE-1 has a small
contribution to the decoherence, and the CCE-2 almost produces the converged results. This means the dominant
decoherence mechanism under weak magnetic field is the nuclear pair correlations. 
Higher order correlations of larger nuclear spin clusters, e.g., the CCE-5 and CCE-6, provide only small corrections in
details.

We further categorize all the nuclear spin pairs into two classes according to the
inter-nuclei distance. The first kind is the nuclear spin dimers. For natural abundance $^{13}$C, there are 
on average about 12 dimers in the bath within 4~nm around the NV center. 
The second class of pairs consists of a large number of separated spin pairs, which have the inter-spin
separations larger than the C-C bond length. In Fig.~\ref{FIG:DDzeroB2}(b), the different effects of these two classes of pairs are shown. 
The decoherence in the weak field regime is dominated by the nuclear dimers [see
Fig.~\ref{FIG:fig3_nonSecular_data}(a) for decoherence by a typical dimer]. 
The separated pairs, which have much slower dynamics than the dimers because of the weaker dipolar interaction, 
only give a small contribution to the total decoherence [see Fig.~\ref{FIG:fig3_nonSecular_data}(b) for decoherence
induced by a typical separated pair]. 

In Fig.~\ref{FIG:DDzeroB2}(c), the roles of different types of interaction under DD control are compared.
Similar to the Hahn echo case, the anisotropic terms of hyperfine coupling and the non-secular part of the
nuclear spin dipolar interaction are the relevant decoherence mechanisms under DD controls in the weak field
regime. 
The secular flip-flop process contributes only a small amount of decoherence [see the blue dashed line in
Fig.~\ref{FIG:DDzeroB2}(c)].
This is consistent with the discussion in Section~\ref{subsubsect:nonsecular}.

\subsubsection{Strong magnetic field regime}
\label{SubsubSect:StrongMag} 
\begin{figure}[tb]
  \includegraphics[width= \columnwidth]{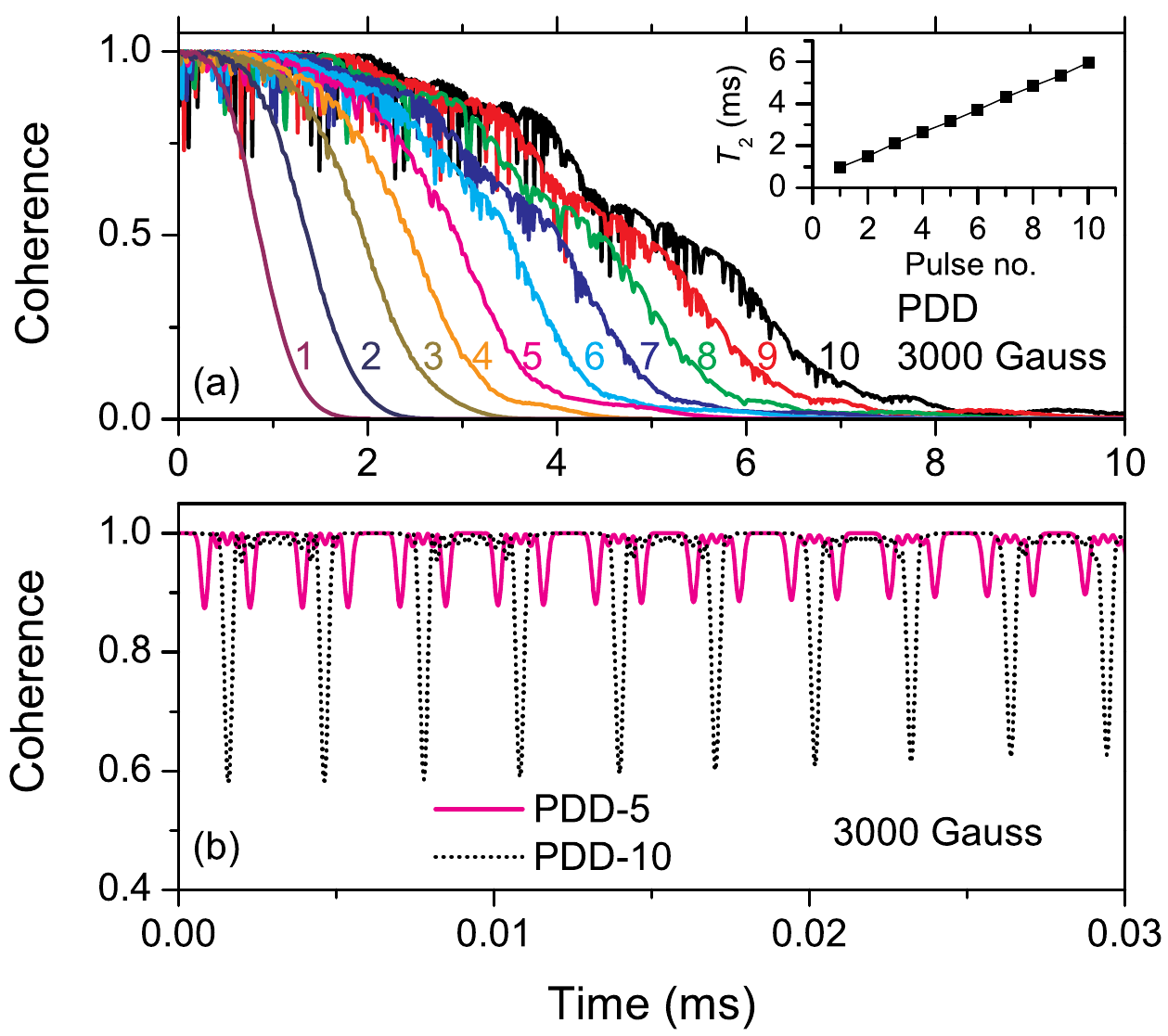}
  \caption{(a) Electron spin coherence between $|0\rangle$ and $|+1\rangle$ under strong magnetic field $B=3000$~Gauss
  and PDD controls. The integer numbers associated with each curves are the pulse numbers of the DD controls.
  Inset: the coherence time $T_2$ linearly increases as increasing the pulse number.   
  (b) The close-up of the initial stage of the decoherence under PDD-5 and PDD-10 controls. Sharp coherence dips are
  caused by the noise from the nuclear spins close to NV center.}
  \label{FIG:fig10_DDhighB}
\end{figure}

Finally, we study the decoherence in the strong magnetic field regime. The coherence under multi-pulse PDD controls
is shown in Fig~\ref{FIG:fig10_DDhighB}(a). 
The overall coherence time $T_2$ is linearly increased with increasing the pulse number [see the inset of
Fig~\ref{FIG:fig10_DDhighB}(a)]. Meanwhile, in the same strong field (e.g., $B=3000$~Gauss), sharp dips appear on
the smooth background decay, and the dips get more profound as
increasing the control pulse number [see Fig.~\ref{FIG:fig10_DDhighB}(b)].

These coherence dips under higher order DD control come from the noises produced by the single spin precession.
Although most of the nuclear spins are frozen by the large Zeeman energy in the strong magnetic field, a few
$^{13}$C spins located close to the NV center (with the hyperfine coupling $\sim 100$~kHz) could produce weak noises to
the NV center spin. In the Hahn echo case, such weak noises can hardly cause any notable effect on the coherence. While in the multi-pulse
PDD cases, the weak noises can be greatly amplified by the pulse sequence, and cause profound coherence dips. 
Under PDD controls, the coherence dips are quasi-periodic. The period is proportional to the pulse number $N$, and the
dip depth is proportional to $N^2$ [see Fig.~\ref{FIG:fig10_DDhighB}(b) and Ref.~\onlinecite{Zhao2011a}].  
This weak noise amplification by multi-pulse DD control has the potential application in the single nuclear spin
detection.\cite{Zhao2011a} The detailed discussion on this prospect will be presented elsewhere.

\begin{figure}[tb]
  \includegraphics[width= \columnwidth]{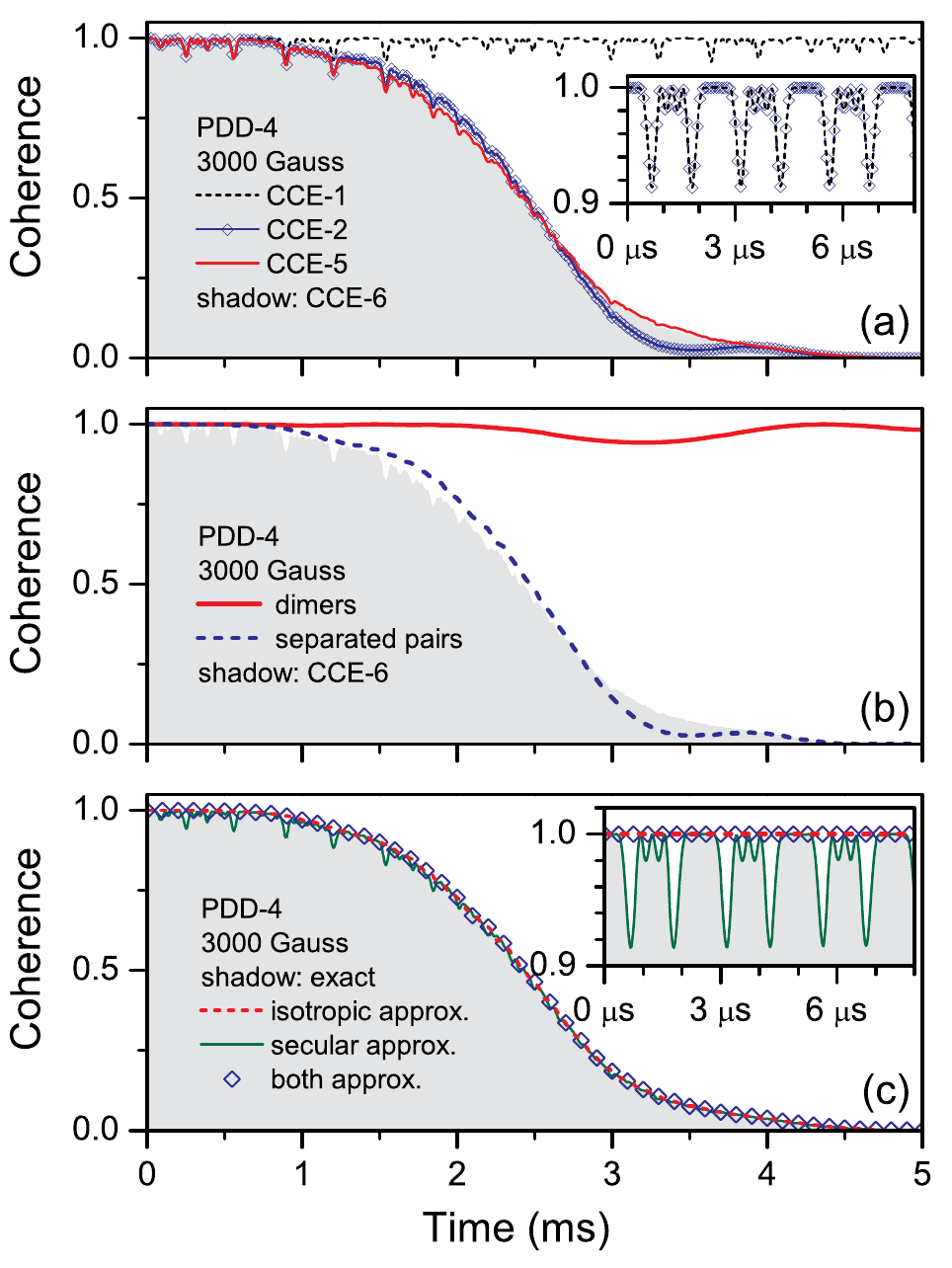}
  \caption{(a) The CCE convergence for PDD-4 under strong magnetic field $B=3000$~Gauss. 
  The 2-spin clusters give the main contribution (CCE-2), and the result converges by including up to 6-spin clusters
  (CCE-6). 
  Inset: close-up for the CCE-1 and CCE-2 contributions to the coherence dips in short time.
  (b) Contributions to the decoherence from $\sim 12$ nuclear spin dimers (red solid line) and the large number of
  separated pairs (blue dashed line) in the bath. The electron spin decoherence under strong magnetic field is mainly
  caused by the separated pairs. (c) Decoherence caused by different types of interaction under $B=3000$~Gauss and
  PDD-4 control. 
  Inset: close-up for the contributions to the coherence dips of different interactions.}
  \label{FIG:DDhighB}
\end{figure}

The origin of the coherence dips is confirmed by the CCE convergence check. As shown in Fig~\ref{FIG:DDhighB}(a),
under PDD-4 control, the single spin dynamics (CCE-1) does not cause the coherence decay but provides the coherence
dips. Similar to the weak field regime, the decoherence is mainly caused by the pair correlations (CCE-2), while
the higher order spin correlations induce corrections to the details.

Furthermore, we compare the contributions of the dimers and the separated pairs. 
As we have shown in our previous study,\cite{Zhao2011a} a dimer located close enough to the NV center (with distance
$\lesssim 1.5$~nm) can induce strong oscillations of the electron spin coherence.
In this paper, we are mainly interested in the overall decoherence, which is
contributed by many weakly coupled pairs in the bath. 
As a result, in Fig.~\ref{FIG:DDhighB}, we choose a bath spin configure without a strongly coupled dimer within
$1.5$~nm to NV center (about $50\%$ probability to find such kind of bath spin configurations in diamond with
natural isotope abundance) to get relative smooth background decoherence profile. 
As shown in Fig~\ref{FIG:DDhighB}(b), in contrast to the weak field regime, the decoherence under strong magnetic
field is dominated by the separated pairs.

The different roles of dimers in decoherence under weak and strong magnetic fields can be understood as follows. 
Because of the energy conservation, only the secular spin flip-flop process is effective in strong magnetic field.
As discussed in Section~\ref{Sect:SpinPair}, the secular spin flip-flop process is associated with the hyperfine field
difference between the two nuclear spins. Because of the small hyperfine field differences for dimers far from the
NV center, the two pseudo-fields $\mathbf{h}_{jk}^{(m)}$ are approximately parallel with each other. 
Thus, according to the pseudo-spin picture in Fig.~\ref{FIG:fig4_BlochSphere}, the remote dimers are not important for
the decoherence. 
This discussion is supported by the analysis of the contributions of different type of interaction shown in
Fig.~\ref{FIG:DDhighB}(c). Except the coherence dips induced by the anisotropic terms [see the inset of
Fig.~\ref{FIG:DDhighB}(c)], neglecting the energy non-conserving spin flipping processes does not change the NV center
spin decoherence behavior. Thus, we conclude that the secular flip-flop of the separated nuclear spin pairs is the main decoherence mechanism in the strong magnetic field regime.

\section{Conclusion}
\label{Sect:conclusion}

In this paper, we systematically investigate the electron spin decoherence of NV centers in $^{13}$C nuclear
spin baths under DD controls. 
We present the different decoherence behaviors in weak, medium and strong magnetic field regimes, and examine the
underlying microscopic mechanisms.  
In medium magnetic field regime, the single nuclear spin precession process induces the coherence collapse and revival
effect. In the weak and strong magnetic field regimes, the decoherence is dominated by two kinds of nuclear spin pair
processes, namely, the non-secular spin flipping of dimers in the weak field regime and the secular spin flip-flops of
the separated nuclear spin pairs in the strong field regime.

The NV center systems investigated here are different from QDs and shallow donors by the dipolar nature
of the hyperfine interaction and the diluteness of the nuclear spin bath. 
The NV center electron spin in the nuclear spin bath is a representative system for deep-level defects. 
The decoherence behavior of NV centers discussed here can be generalized to other similar systems.\cite{Weber2010}
Better understanding of the underlying decoherence mechanisms will benefit various applications of such kind of systems,
such as quantum information processing\cite{Wrachtrup2001,Wrachtrup2006,Childress2006a,Dutt2007} and ultra-sensitive
metrology.\cite{Maze2008,Balasubramanian2008,Taylor2008,Hall2010,Hall2009,Zhao2011a,Dolde2011,DeLange2011}

\section*{ACKNOWLEDGEMENTS}
This work was supported by Hong Kong RGC/GRF (CUHK402208, CUHK402209, and CUHK402410), CUHK Focused Investments
Scheme, and National Natural Science Foundation of China Project 11028510.

\appendix*
\section{Single-spin dynamics under PDD controls}
\label{Sect:App}

\begin{figure}[tb]
  \includegraphics[width= 0.5\columnwidth]{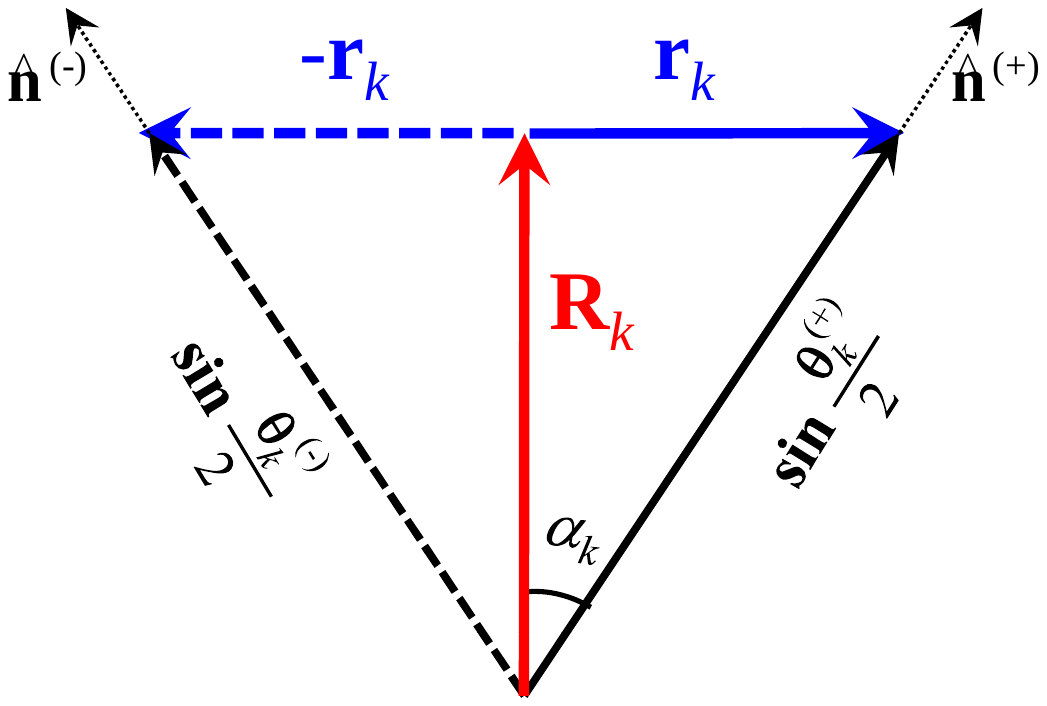}
  \caption{Geometric picture of the effective bath spin precession about the
  axes along $\hat{\mathbf{n}}_k^{(\pm)}$ with precession angles $\theta_k^{(\pm)}$, under PDD-$k$ ($k\ge 1$)
  control. The angle between the precession axes is $2\alpha_k$. 
  The vectors $\sin(\boldsymbol{\theta}_k^{(\pm)}/2)$ are decomposed into the common part $\mathbf{R}_k$ and the
  difference part $\mathbf{r}_k$.
  \label{FIG:Appendix}
  }
\end{figure}

Here we present the derivation of the electron spin decoherence induced by single spin dynamics under PDD controls.
To be specific, we will use the terminology for single-spin dynamics, but the discussion automatically applies to the
pseudo-spin dynamics. 
Given the electron spin in state $|m\rangle$, the free evolution operator of a nuclear spin is denoted as
\begin{equation}
U_0^{(m)}(\tau)=e^{-i\boldsymbol{\theta}_0^{(m)}\cdot\boldsymbol{\sigma}/2}=\cos\frac{\theta_0^{(m)}}{2}
-i\boldsymbol{\sigma}\cdot\sin\frac{\boldsymbol{\theta}_0^{(m)}}{2},
\end{equation}
where $\sin(\mathbf{v})\equiv (\mathbf{v}/v)\sin(v)$, and the angle
$\boldsymbol{\theta}_0^{(m)}=\gamma_{\text{n}}\mathbf{h}^{(m)}\tau$ is the precession angle for the nuclear spin about
the effective field $\mathbf{h}^{(m)}$ [see Eq.~(\ref{Eq:SingleSpinHami})]. Here we use the integer subscript $k$ to
indicate the DD control order ($k=0$ means free evolution, and 1 for Hahn echo), while the nuclear spin index $j$ is
omitted for the sake of clarity. According to Eq.~(\ref{Eq:CCE_Lc}), the single nuclear spin contribution to Hahn echo is
\begin{eqnarray}
\label{Eq:AppendixHahn}
L_{\text{Hahn}}\left(2\tau\right)&=&\frac{1}{2}\text{Tr}\left[\left(U_0^{(0)}(\tau)U_0^{(+1)}(\tau)\right)^{\dagger}U_0^{(+1)}(\tau)U_0^{(0)}(\tau)\right]\notag\\
&=&\frac{1}{2}\text{Tr}\left[\left(U_1^{(+)}(\tau)\right)^{\dagger}U_1^{(-)}(\tau)\right].
\end{eqnarray} 
Here we have defined the effective evolution operators $U_1^{(\pm)}(\tau)$ under the Hahn echo control as
\begin{eqnarray}
\label{Eq:AppendixU1}
U_1^{(\pm)}(\tau)&=&U_0^{(0/+1)}(\tau)U_0^{(+1/0)}(\tau)\equiv e^{-i\boldsymbol{\theta}_1^{(\pm)} \cdot
\boldsymbol{\sigma}/2}\notag\\
&=&\cos\frac{\theta_1}{2}-i\boldsymbol{\sigma}\cdot\sin\frac{\boldsymbol{\theta}_1^{(\pm)}}{2},
\end{eqnarray} 
where the effective precession angle $\boldsymbol{\theta}_1^{(\pm)} \equiv \theta_1\hat{\mathbf{n}}_1^{(\pm)}$ with
$\hat{\mathbf{n}}_1^{(\pm)}$ being the directions of the effective precession axis and the magnitude $\theta_1$ is
determined by\cite{Liu2007}
\begin{equation}
\cos\frac{\theta_1}{2}=\cos\frac{\theta_0^{(+1)}}{2}\cos\frac{\theta_0^{(0)}}{2}
-\sin\frac{\boldsymbol{\theta}_0^{(+1)}}{2}\cdot\sin\frac{\boldsymbol{\theta}_0^{(0)}}{2}.
\end{equation}
The vector of effective precession angles are decomposed into the common part $\mathbf{R}_1$ and the difference part
$\mathbf{r}_1$ as
\begin{eqnarray}
\label{Eq:HahnPrecession}
\sin\frac{\boldsymbol{\theta}_1^{(\pm)}}{2}&=&\mathbf{R}_1\pm\mathbf{r}_1,
\end{eqnarray}
with 
\begin{subequations}
\label{Eq:HahnPrecessionRr}
\begin{eqnarray}
\mathbf{R}_1&=&\cos\frac{\theta_0^{(+1)}}{2}\sin\frac{\boldsymbol{\theta}_0^{(0)}}{2}
+\cos\frac{\theta_0^{(0)}}{2}\sin\frac{\boldsymbol{\theta}_0^{(+1)}}{2},\\
\mathbf{r}_1&=&\sin\frac{\boldsymbol{\theta}_0^{(0)}}{2}\times\sin\frac{\boldsymbol{\theta}_0^{(+1)}}{2}.
\end{eqnarray}
\end{subequations}
The geometric picture of the effective precession is shown in Fig.(\ref{FIG:Appendix}).
Notice that the common part $\mathbf{R}_1$ and the difference part $\mathbf{r}_1$ are perpendicular to each other.
The angle between the two effective precession axes is determined by 
\begin{equation}
\tan\alpha_1=\frac{|\mathbf{r}_1|}{|\mathbf{R}_1|}.
\end{equation}
substituting Eqs.~(\ref{Eq:AppendixU1})-(\ref{Eq:HahnPrecessionRr}) into Eq.~(\ref{Eq:AppendixHahn}),
one obtains 
\begin{eqnarray}
L_{\text{Hahn}}\left(2\tau\right)&=&1-2|\mathbf{r}_1|^2\notag\\
&=&1-2\sin^2\varphi\sin^2\frac{\theta_0^{(0)}}{2}\sin^2\frac{\theta_0^{(+1)}}{2},
\label{Eq:AppendixHahnResult}
\end{eqnarray}
where $\varphi$ is the angle between the effective fields $\mathbf{h}^{(0)}$ and $\mathbf{h}^{(+1)}$.

The effective evolution operators for higher order PDD controls are defined by combining the evolution operators of
lower orders in the similar way. 
For the even order PDD control PDD-$2k$, the effective evolution operators $U_{2k}^{(\pm)}(\tau)$ are
\begin{eqnarray}
\label{Eq:PDD2kEVO}
U_{2k}^{(\pm)}(\tau)&=&\left[U_1^{(\mp)}(\tau)U_1^{(\pm)}(\tau)\right]^k\equiv \left[e^{-i\boldsymbol{\theta}_2^{(\pm)}
\cdot \boldsymbol{\sigma}/2}\right]^k\notag\\
&=&\cos\frac{\theta_{2k}}{2}-i\boldsymbol{\sigma}\cdot\sin\frac{\boldsymbol{\theta}_{2k}^{(\pm)}}{2},
\end{eqnarray} 
where
$\boldsymbol{\theta}_{2k}^{(\pm)}=\theta_{2k}\hat{\mathbf{n}}_{2k}^{(\pm)}=k\boldsymbol{\theta}_2^{(\pm)}=k\theta_2\hat{\mathbf{n}}_2^{(\pm)}$.
The precession angle $\theta_2$ can be obtained through
\begin{subequations}
\begin{eqnarray}
\label{Eq:HahnPrecession2}
\cos\frac{\theta_2}{2}&=&\cos^2\frac{\theta_1}{2}-
\sin\frac{\boldsymbol{\theta}_1^{(+)}}{2}\cdot\sin\frac{\boldsymbol{\theta}_1^{(-)}}{2}\\
&=&\cos\theta_0^{(+)}\cos\theta_0^{(-)}-\sin\boldsymbol{\theta}_0^{(+)}\cdot\sin\boldsymbol{\theta}_0^{(-)}.
\end{eqnarray}
\end{subequations}
The directions of precession axes are determined by the vector $\sin(\boldsymbol{\theta}_{2k}^{(\pm)}/2)$, which
are decomposed into the common and difference parts in the similar way of Eq.~(\ref{Eq:HahnPrecession}) as
\begin{eqnarray}
\sin\frac{\boldsymbol{\theta}_{2k}^{(\pm)}}{2}&=&\mathbf{R}_{2k}\pm\mathbf{r}_{2k}.
\end{eqnarray}
For $k=1$, the vectors $\mathbf{R}_2$ and $\mathbf{r}_2$ are calculated as
\begin{subequations}
\begin{eqnarray}
\mathbf{R}_2&=&2 \cos\frac{\theta_1}{2} \mathbf{R}_1,\\
\mathbf{r}_2&=&2\mathbf{R}_1\times\mathbf{r}_1.
\end{eqnarray}
\end{subequations}
For $k>1$, the vectors $\sin(\boldsymbol{\theta}_{2k}^{(\pm)}/2)$ have the same direction as
$\sin(\boldsymbol{\theta}_{2}^{(\pm)}/2)$, but with a magnitude $\sin(k\theta_2/2)$.
The angle between
effective precession axes is determined by
\begin{equation}
\label{Eq:PDD2angle}
\tan\alpha_2=\frac{|\mathbf{r}_2|}{|\mathbf{R}_2|}=\frac{|\mathbf{r}_1|}{\cos\theta_1/2},
\end{equation} 
which is the result shown in Eq.~(\ref{Eq:PDDanglesA}).

With Eqs.~(\ref{Eq:PDD2kEVO})-(\ref{Eq:PDD2angle}), the coherence under PDD-$2k$ control is expressed in terms of the
angles $\theta_2$ and $\alpha_2$ as
\begin{subequations}
\begin{eqnarray}
&&L_{2k}(4k\tau)=\frac{1}{2}\text{Tr}\left[\left(U_{2k}^{(-)}(\tau)\right)^{\dagger}U_{2k}^{(+)}(\tau)\right]\\
&=&1-2|\mathbf{r}_{2k}|^2=1-2\sin^2\alpha_2\sin^2\frac{k\theta_2}{2}.
\end{eqnarray}
\end{subequations}

The odd order PDD control PDD-$(2k+1)$ is a combination of PDD-$2k$ and a Hahn echo. The evolution
operator is $U_{2k+1}^{(\pm)}=U_1^{(\pm)}(\tau)U_{2k}^{(\pm)}(\tau)$, and the coherence at $t=(4k+2)\tau$ is
\begin{subequations}
\begin{eqnarray}
L_{2k+1}(t)&=&\frac{1}{2}\text{Tr}\left[\left(U_{2k+1}^{(+)}(\tau)\right)^{\dagger}U_{2k+1}^{(-)}(\tau)\right]\\
&=&\frac{1}{2}\text{Tr}\left[Q_1^{\dagger}(\tau)Q_{2k}(\tau)\right],
\end{eqnarray}
\end{subequations} 
where 
\begin{subequations}
\begin{align}
&&Q_1(\tau)=\left(U_1^{(-)}(\tau)\right)^{\dagger}U_1^{(+)}(\tau)=L_{\text{Hahn}}(2\tau)-i\boldsymbol{\sigma}\cdot\boldsymbol{\xi}_1,\\
&&Q_{2k}(\tau)=U_{2k}^{(-)}(\tau)\left(U_{2k}^{(+)}(\tau)\right)^{\dagger}=L_{2k}(4k\tau)-i\boldsymbol{\sigma}\cdot\boldsymbol{\xi}_{2k},
\end{align}
\end{subequations}
with the traceless terms determined by the vectors
\begin{subequations}
\begin{eqnarray}
\boldsymbol{\xi}_1&=&2\left(\mathbf{r}_1\times\mathbf{R}_1+\cos\frac{\theta_1}{2}\mathbf{r}_1\right),\\
\boldsymbol{\xi}_{2k}&=&2\left(\mathbf{r}_{2k}\times\mathbf{R}_{2k}-\cos\frac{\theta_{2k}}{2}\mathbf{r}_{2k}\right).
\end{eqnarray}
\end{subequations}
Then, the coherence $L_{2k+1}(t)$ is the sum of $L_{\text{Hahn}}(2\tau)L_{2k}(4k\tau)$ and the correction
term
\begin{eqnarray}
L_{\text{corr}}(\tau)&=&\boldsymbol{\xi}_1\cdot\boldsymbol{\xi}_{2k}.
\end{eqnarray}

%
\end{document}